\documentclass{aa}
\usepackage{rotating}
\usepackage{graphicx}
\usepackage{txfonts}
\begin{document}
\title{Simultaneous X-ray and optical spectroscopy\newline 
of the Oef supergiant $\lambda$~Cep\thanks{Based on observations collected with {\it XMM-Newton}, an ESA science mission with instruments and contributions directly funded by ESA member states and the USA (NASA), and with the TIGRE telescope (La Luz, Mexico) and the 1.5\,m telescope at Observatoire de Haute Provence (France).}}
\author{G.\ Rauw\inst{1} \and A.\ Herv\'e\inst{2} \and Y.\ Naz\'e\inst{1}\thanks{Research Associate FRS-FNRS (Belgium)} \and J.N.\ Gonz\'alez-P\'erez\inst{3} \and A.\ Hempelmann\inst{3} \and M.\ Mittag\inst{3} \and J.H.M.M.\ Schmitt\inst{3} \and \newline K.-P.\ Schr\"oder\inst{4} \and E.\ Gosset\inst{1}\thanks{Senior Research Associate FRS-FNRS (Belgium)} \and P.\ Eenens\inst{4} \and J.M.\ Uuh-Sonda\inst{4}}
\offprints{G.\ Rauw}
\mail{rauw@astro.ulg.ac.be}
\institute{Groupe d'Astrophysique des Hautes Energies, Institut d'Astrophysique et de G\'eophysique, Universit\'e de Li\`ege, Quartier Agora, All\'ee du 6 Ao\^ut, 19c, B\^at B5c, 4000 Li\`ege, Belgium
\and LUPM, Universit\'e de Montpellier 2, CNRS, Place Eug\`ene Bataillon, 34095 Montpellier, France
\and Hamburger Sternwarte, Universit\"at Hamburg, Gojenbergsweg 112, 21029 Hamburg, Germany
\and Departamento de Astronom\'{\i}a, Universidad de Guanajuato, Apartado 144, 36000 Guanajuato, GTO, Mexico}
\abstract{Probing the structures of stellar winds is of prime importance for the understanding of massive stars. Based on their optical spectral morphology and variability, the stars of the Oef class have been suggested to feature large-scale structures in their wind.}{High-resolution X-ray spectroscopy and time-series of X-ray observations of presumably-single O-type stars can help us understand the physics of their stellar winds.}{We have collected XMM-Newton observations and coordinated optical spectroscopy of the O6\,Ief star $\lambda$~Cep to study its X-ray and optical variability and to analyse its high-resolution X-ray spectrum. We investigate the line profile variability of the \ion{He}{ii} $\lambda$\,4686 and H$\alpha$ emission lines in our time series of optical spectra, including a search for periodicities. We further discuss the variability of the broadband X-ray flux and analyse the high-resolution spectrum of $\lambda$~Cep using line-by-line fits as well as a code designed to fit the full high-resolution X-ray spectrum consistently.}{During our observing campaign, the \ion{He}{ii} $\lambda$\,4686 line varies on a timescale of $\sim 18$\,hours. On the contrary, the H$\alpha$ line profile displays a modulation on a timescale of 4.1\,days which is likely the rotation period of the star. The X-ray flux varies on time-scales of days and could in fact be modulated by the same 4.1\,days period as H$\alpha$, although both variations are shifted in phase. The high-resolution X-ray spectrum reveals broad and skewed emission lines as expected for the X-ray emission from a distribution of wind-embedded shocks. Most of the X-ray emission arises within less than 2\,$R_*$ above the photosphere.}{The properties of the X-ray emission of $\lambda$~Cep generally agree with the expectations of the wind-embedded shock model. There is mounting evidence for the existence of large-scale structures that modulate the H$\alpha$ line and about 10\% of the X-ray emission of $\lambda$~Cep.}
\authorrunning{Rauw et al.}
\titlerunning{X-ray spectroscopy of $\lambda$~Cep}
\maketitle

\section{Introduction}\label{intro}
High-resolution X-ray spectroscopy with the RGS and the HETG instruments onboard of {\it XMM-Newton} and {\it Chandra}, respectively, has revolutionized our view of the X-ray emission of massive stars. One of the first O-type stars that was observed in X-rays at high spectral resolution was the O4\,Ief supergiant $\zeta$~Pup (Cassinelli et al.\ \cite{Cassinelli}, Kahn et al.\ \cite{Kahn}), often considered as the prototype O-supergiant. The spectrum revealed broad, skewed, and blue-shifted lines, as expected for an X-ray emitting plasma distributed throughout the stellar wind and moving along with this wind (MacFarlane et al.\ \cite{MacFarlane}, Owocki \& Cohen \cite{OC1}). However, a detailed analysis of the shape of the spectral lines indicated that there is less absorption in the wind than expected given the mass loss rate of the star inferred under the assumption of a homogeneous wind. Moreover, observations of a small sample of other, relatively nearby and hence X-ray bright, single O-type stars with {\it XMM-Newton} and {\it Chandra} revealed lines that were less broad, less skewed and less blue-shifted than expected (see G\"udel \& Naz\'e \cite{GN} and references therein). These findings were interpreted as evidence for the winds of O-type stars being clumpy (or even porous) and the mass-loss rates being lower than previously thought (Feldmeier et al.\ \cite{Feldmeier}, Owocki \& Cohen \cite{OC2}, Oskinova et al.\ \cite{Oskinova},  Cohen et al.\ \cite{Cohen,Cohen2}). 

$\zeta$~Pup displays X-ray line morphologies as predicted for a wind-embedded plasma, but has several properties that make it stand out from the sample of presumably single O-stars observed at high-resolution in X-rays. For instance, it has the largest mass-loss rate and one of the earliest spectral types among the stars of this sample. Moreover, its optical spectrum displays peculiarities that are partly reflected by its Oef spectral classification. 

In this context, we decided to observe another Oef star, as similar as possible to $\zeta$~Pup. The best candidate for this purpose was the O6\,Ief supergiant $\lambda$~Cep (= HD~210\,839). The peculiarities of the optical spectrum of $\lambda$~Cep and its similarities with $\zeta$~Pup were already noted by Wilhelmina Fleming (Pickering \& Fleming \cite{PF}). In the early 1970's, it was found that both $\lambda$~Cep and $\zeta$~Pup belong to a handful of stars that display a double-peaked \ion{He}{ii} $\lambda$\,4686 emission line in their spectrum. These objects were classified as so-called Onfp stars by Walborn (\cite{Walborn}) or Oef stars by Conti \& Leep (\cite{CL}). $\zeta$~Pup ($V = 2.25$) and $\lambda$~Cep ($V = 5.05$) are by far the brightest members of this class (both in optical and X-rays). In addition, both $\zeta$~Pup and $\lambda$~Cep are most-probably single runaway stars (Gies \cite{Gies}) and are rather fast rotators (Conti \& Ebbets \cite{CE}, Penny \cite{Penny}, Howarth et al.\ \cite{Howarth}, Sim\'on-D\'{\i}az \& Herrero \cite{SDH1,SDH}). The peculiar morphology of the \ion{He}{ii} $\lambda$\,4686 line of Oef stars is likely related to this fast rotation. Indeed, Hillier et al.\ (\cite{Hillier}) demonstrated that the double-peaked profile of this line stems from a complex interplay of emission and absorption of a rotating wind\footnote{Hillier et al.\ (\cite{Hillier}) assume a rigidly rotating wind out to a distance where the wind velocity reaches 20\,km\,s$^{-1}$, and conservation of angular momentum beyond that point.}. 
Hillier et al.\ (\cite{Hillier}) further showed that the box-shaped morphology of the \ion{N}{iii} $\lambda\lambda$\,4634-4642 blend is also due to the effect of rotation, but this time on a purely photospheric emission line. Harries (\cite{Harries}), Harries et al.\ (\cite{Harries2}) and Vink et al.\ (\cite{Vink}) reported complex linear polarization effects across the H$\alpha$ line profiles of $\zeta$~Pup and $\lambda$~Cep. These authors attributed these effects to the impact of the rapid rotation of the stars on their winds. 

The idea of a corotating wind coupled with presumably rotational modulations of the optical emission lines of $\zeta$~Pup led Moffat \& Michaud (\cite{MM}) to the suggestion that the winds of Oef stars might be shaped by a dipolar stellar magnetic field. However, David-Uraz et al.\ (\cite{DU}) recently revisited the magnetic properties of a sample of OB stars, including $\zeta$~Pup and $\lambda$~Cep. 
Their goal was to establish whether or not the discrete absorption components (DACs) seen in the UV spectra of their sample stars are associated with large-scale dipolar fields. No significant longitudinal magnetic field was observed in any of the stars: for $\lambda$~Cep an upper limit on the dipolar field strength of 136\,G was inferred, corresponding to a wind confinement parameter $\eta_* = \frac{B^2_{\rm eq}\,R^2_*}{\dot{M}\,v_{\infty}} \leq 0.15$ which should have no impact on the global geometry of the wind of $\lambda$~Cep nor on its X-ray emission.

In this paper, we present the very first analysis of the X-ray spectrum and lightcurve of $\lambda$~Cep, as well as the results of a simultaneous monitoring of the star in optical spectroscopy. Our observations are presented in Sect.\,\ref{sectobs}. The variability of $\lambda$~Cep, in the optical and the X-ray domain, is addressed in Sects.\,\ref{sectvar} and \ref{varX}, respectively. Sect.\,\ref{sectX} presents our analysis of the X-ray spectrum of $\lambda$~Cep. Finally, Sect.\,\ref{sectdisc} discusses our results and presents our conclusions.

\section{Observations \label{sectobs}}
We collected four observations of $\lambda$~Cep with {\it XMM-Newton} (Jansen et al.\ \cite{Jansen}). Each of these observations had a nominal duration of 70 - 95\,ks (see Table\,\ref{journalX}). Observations I and III were affected by a rising background level towards the end of the observation (typically during the last 10 - 15\% of the exposure). The corresponding time intervals were discarded.  

The EPIC instruments (Turner et al.\ \cite{MOS}, Str\"uder et al.\ \cite{pn}) were operated in full frame mode and with the thick filter to reject optical and UV photons. The thick filters efficiently suppress the optical contamination for sources with $V > 4$ (MOS 1 \& 2) and $V> 1$ (pn). Given the optical magnitude of $\lambda$~Cep ($V=5.05$) we thus expect no optical loading of the EPIC CCDs. 
The reflection grating spectrometer (RGS, den Herder et al.\ \cite{RGS}) was operated in normal spectroscopy mode. 

The raw data were processed with the Scientific Analysis System (SAS) software version 13.0. The EPIC spectra of the source were extracted over a circular region centered on the 2\,MASS coordinates of $\lambda$~Cep and adopting an extraction radius of 50\,arcsec. The background spectrum was evaluated from a nearby, source-free circular region located on the same detector chip as the source itself. The RGS1 and RGS2 first and second order spectra were extracted for each observation individually, and were then combined for the subsequent analysis. Since $\lambda$~Cep is a rather isolated X-ray source, there is no risk of contamination of its RGS spectrum by nearby bright X-ray emitters. Lightcurves of the source and the background were extracted for each EPIC instrument and converted into equivalent on-axis, full-PSF count rates using the {\it epiclccorr} command under SAS. We built lightcurves over the full energy band and for three different time bins (100\,s, 500\,s and 1\,ks), as well as over soft (0.2 -- 1.0\,keV) and hard (1.0 -- 10.0\,keV) bands with 1\,ks time bins.

\begin{table}
\caption{Journal of the {\it XMM-Newton} observations \label{journalX}}
\begin{center}
\begin{tabular}{c c c}
\hline
Obs & Date          & Exposure time \\
    & JD$-$2\,450\,000  &     (ks)      \\
\hline
I   & 6456.677      &      82.5     \\
II  & 6508.319      &      75.8     \\
III & 6510.473      &      94.9     \\
IV  & 6514.317      &      70.7     \\
\hline
\end{tabular}
\tablefoot{The date of the observation is given at mid-exposure. The exposure time indicated in the third column corresponds to the RGS1 instrument.}
\end{center}
\end{table}

In support of the X-ray observations, we collected optical spectroscopy. For this purpose, we used the AURELIE spectrograph (Gillet et al.\ \cite{Gillet}) at the 1.52\,m telescope of the Observatoire de Haute Provence (OHP, France) and the refurbished HEROS spectrograph at the TIGRE telescope (Schmitt et al.\ \cite{Schmitt}) at La Luz Observatory (Guanajuato, Mexico). The AURELIE spectra have a resolving power of 7000 and cover the region between 4450 and 4880\,\AA. The exposure times were 10\,min per observation and the mean S/N ratio was 340. Between two and four consecutive exposures were taken per night. The HEROS echelle spectra have a resolving power of 20\,000 and cover the full optical range, though with a small gap near 5800\,\AA. The integration times were 5\,min, resulting in a mean S/N of 160 per pixel. The spectra were usually taken by series of three consecutive integrations. Whenever possible, several series were taken during the night (beginning, middle and end of the night). 

\begin{figure}
\resizebox{8cm}{!}{\includegraphics{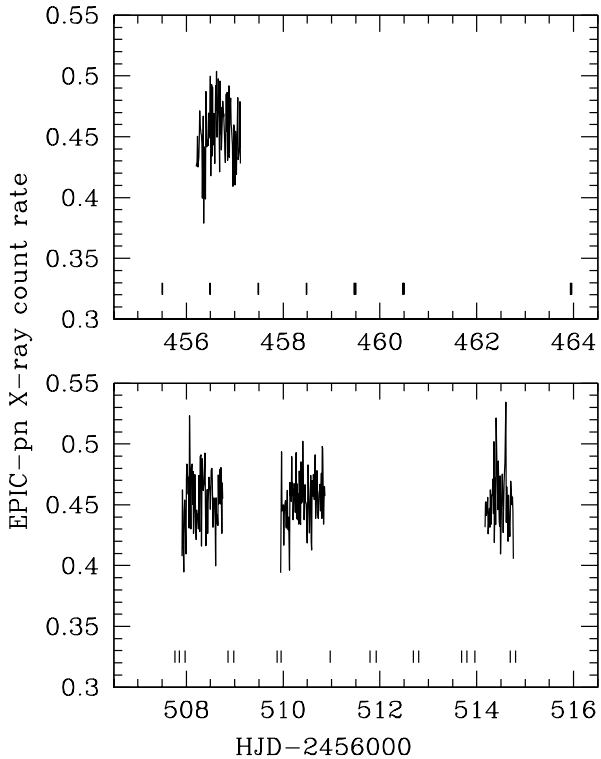}}
\caption{The EPIC-pn lightcurve of $\lambda$~Cep is shown along with tickmarks that indicate the times of the AURELIE and HEROS spectra. The top and bottom panels correspond to the June and August 2013 campaigns, respectively.\label{montage}}
\end{figure}

A first set of optical spectra was taken in June 2013 (16 AURELIE spectra and 3 HEROS spectra) around the time of X-ray observation I. A more intensive monitoring was performed in August 2013 covering the X-ray observations II to IV (35 HEROS spectra). Figure\,\ref{montage} illustrates the EPIC-pn lightcurve of $\lambda$~Cep along with the sampling of the optical spectra. 

The HEROS data were reduced with the corresponding reduction pipeline (Mittag et al.\ \cite{Mittag}, Schmitt et al.\ \cite{Schmitt}) whilst the AURELIE spectra were processed with the MIDAS software provided by ESO. The spectra were normalized self-consistently using a series of continuum windows.\\

Unless otherwise stated, the quoted error bars in our analysis of the optical and X-ray data refer to 1$\sigma$ errors.

\section{Optical Variability \label{sectvar}}
Spectroscopic variability in the optical and UV domain is widespread among O supergiants. This variability can arise from small- and large-scale structures in the stellar winds, non-radial pulsations, or magnetic fields. In the case of $\lambda$~Cep, the variability of the \ion{He}{ii} $\lambda$\,4686 and H$\alpha$ emission lines was first noted by Conti \& Frost (\cite{CF}). The variability of the star's emission lines has been subsequently studied by a number of authors (Lacy \cite{Lacy}, Leep \&  Conti \cite{LC}, Ebbets \cite{Ebbets}, Grady et al.\ \cite{Grady}, McCandliss \cite{McCandliss}, Henrichs et al.\ \cite{Henrichs}, Henrichs \cite{Henrichs2}, Kaper et al.\ \cite{Kaper1,Kaper2}, Markova et al.\ \cite{Markova}). Although some quasi-cyclic behaviour with a typical timescale near 2\,days, interpreted as half the rotation period, was found at some epochs (Henrichs \& Sudnik \cite{HenSud2}), the variations do not seem to be strictly (mono)periodic, but are reminiscent of the epoch-dependent variability found in other Oef stars (Rauw et al.\ \cite{bd+60_2522}, De Becker \& Rauw \cite{Oef})\footnote{Quite recently, Howarth \& Stevens (\cite{2Ian}) reported however on a period of 1.78\,days in spaceborne photometry of $\zeta$~Pup. This period was found to be stable over the three year interval covered by the data and was attributed by Howarth \& Stevens (\cite{2Ian}) to low-$l$ pulsations.}. Henrichs \& Sudnik (\cite{HenSud2}) accordingly suggest that the variations might arise from the presence of several transient short-lived magnetic loops corotating with the star. The localized magnetic fields responsible for these loops could take their origin in sub-surface convective layers of the star.  

Absorption lines were also found to be variable (e.g.\ Fullerton et al.\ \cite{FGB}), and de Jong et al.\ (\cite{dJ}, see also Kholtygin et al.\ \cite{Kholtygin}) claimed the detection of low-order non-radial pulsations (NRPs) with periods of 12.3 ($l = 3$) and 6.6\,hrs ($l = 5$). However, a recent multi-epoch analysis of the absorption line profile variability revealed a more complex situation (Uuh-Sonda et al.\ \cite{Jorge,Jorge2}). The frequency content of the power spectrum of these spectral variations considerably changes from one epoch to the other and no stable frequency was found that could unambiguously be attributed to pulsations. As becomes clear from Fig.\,\ref{montage}, the sampling of our new optical data is not suited to study such rapid variations. 

In the study of our optical time series, we accordingly focus on the longer-term variability exhibited by two emission lines (\ion{He}{ii} $\lambda$\,4686 and H$\alpha$) which are at least partially formed in the wind. The H$\alpha$ line is heavily affected by telluric absorption features that need to be corrected to avoid biasing our variability analysis. As a first step, we thus used the {\tt telluric} tool within IRAF along with the list of telluric lines of Hinkle et al.\ (\cite{Hinkle}) to remove these absorptions in the H$\alpha$ region.

The time-series of continuum-normalized and telluric-corrected spectra were analysed by means of dedicated tools. The spectroscopic variability was first quantified via the temporal variance spectrum (TVS, Fullerton et al.\ \cite{FGB}). In the blue spectral range, the strongest variability affects the \ion{He}{ii} $\lambda$\,4686 emission line. Beside the \ion{He}{i} $\lambda$\,4471, \ion{He}{ii} $\lambda$\,4542, and \ion{N}{iii} $\lambda\lambda$ 4509-4522 absorptions, and the \ion{N}{iii} $\lambda\lambda$\,4634-4642 emission lines discussed by Uuh-Sonda et al.\ (\cite{Jorge}), strong variability is also found in the H$\beta$ absorption line. In the red spectra, the H$\alpha$ emission displays by far the strongest line-profile variability.

To characterize the timescales of the line profile variability, we have then applied a 2D-Fourier analysis (Rauw et al.\ \cite{HD93521}). In this method, the Fourier periodogram is computed at each wavelength bin by means of the algorithm of Heck et al.\ (\cite{HMM}) modified by Gosset et al.\ (\cite{Gosset}). This latter method explicitly accounts for the irregular sampling of the time-series.

\begin{figure}
\resizebox{8cm}{!}{\includegraphics{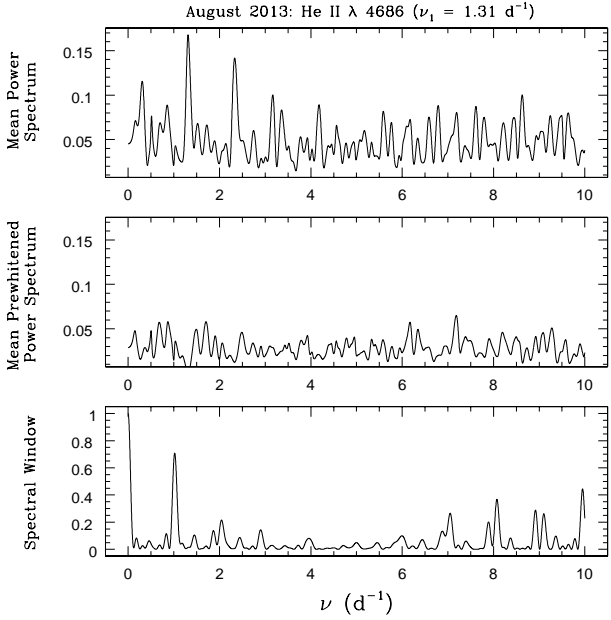}}
\caption{Periodogram of the \ion{He}{ii} $\lambda$\,4686 emission line as observed in August 2013. In the mean periodogram over the wavelength interval 4678 - 4692\,\AA, as computed from the time series of HEROS spectra (top panel), the highest peak is found at the frequency $\nu_1 = 1.315$\,day$^{-1}$. The residual periodogram, after prewhitening the data for the signal corresponding to the $\nu_1$ frequency, is shown in the middle panel. The bottom panel provides the power spectral window of our time series.\label{period4686}}
\end{figure}

\begin{figure}
\resizebox{8cm}{!}{\includegraphics{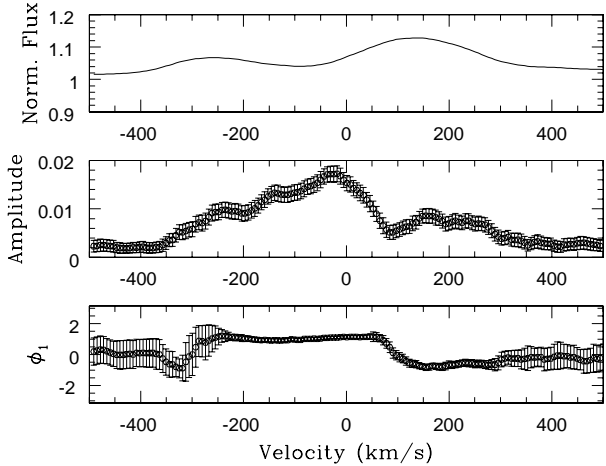}}
\caption{Variations of the \ion{He}{ii} $\lambda$\,4686 emission line with the $\nu_1$ frequency in August 2013. From top to bottom, the panels illustrate the mean line profile, the amplitude and the phase constant of the modulation as a function of wavelength.\label{nu1}}
\end{figure}

For the August 2013 dataset, we find that the periodogram of the \ion{He}{ii} $\lambda$\,4686 line displays a strong peak at $\nu_1 = (1.315 \pm 0.014)$\,day$^{-1}$ (or P = $18.3 \pm 0.2$\,hrs). We have prewhitened the data for the sinusoidal signal corresponding to this frequency. Figure\,\ref{period4686} illustrates the mean periodogram before and after prewhitening. From this figure, one clearly sees that $\nu_1$ accounts for most of the power in the periodogram. The changes of the amplitude and phase constant of the $\nu_1$ modulation as a function of radial velocity are shown in Fig.\,\ref{nu1}. The errors were estimated using Monte-Carlo simulations (see Rauw et al.\ \cite{HD93521}). The modulation extends over the full width of the line, although its amplitude is largest in the central reversal between the two emission peaks.

The other lines in the blue domain that display significant variability, the absorption lines \ion{He}{i} $\lambda$\,4471, \ion{He}{ii} $\lambda$\,4542, and H$\beta$ have distinct periodograms that are dominated by two frequencies, 0.46 and 0.57\,day$^{-1}$, which are aliases of each other. This points towards a different mechanism for the variations of photospheric and wind lines. We stress here that the detected frequencies should not be considered as reflecting strict periodicities. Indeed, $\lambda$~Cep and other Oef stars are known to display variability with timescales that can change quite significantly from one epoch to the other (see above). Moreover, some part of the variability of the emission lines is likely due to statistical fluctuations of the number of small-scale clumps in the wind (see e.g.\ Eversberg et al.\ \cite{Eversberg}) and is thus not expected to follow a periodic behaviour. Our current results should therefore be regarded as indications of the timescale of variations at the epoch of the X-ray observations. 
 
\begin{figure}
\resizebox{8cm}{!}{\includegraphics{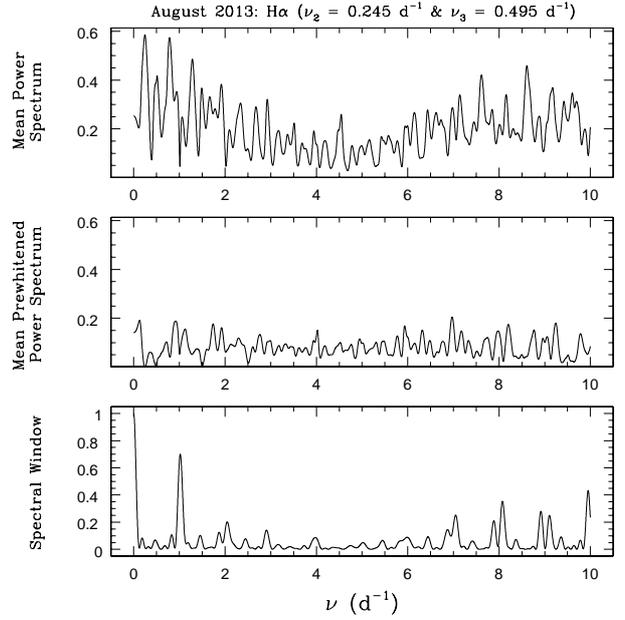}}
\caption{Same as Fig.\,\ref{period4686}, but for the H$\alpha$ emission line as observed in August 2013. The mean periodogram over the wavelength interval 6545 - 6580\,\AA\ (top panel) reveals its highest peak at $\nu_2 = 0.245$\,day$^{-1}$. The middle panel displays the residual periodogram once the data have been prewhitened for the signal corresponding to the $\nu_2$ and $\nu_3$ frequencies. 
\label{periodhalpha}}
\end{figure}

The periodogram of the August 2013 H$\alpha$ data (see Fig.\,\ref{periodhalpha}) reveals no evidence for modulation at the  $\nu_1$ frequency. Instead, most of the power in the periodogram is due to two frequencies $\nu_2 = (0.245 \pm 0.014)$\,day$^{-1}$ and $\nu_3 = (0.495 \pm 0.014)$\,day$^{-1}$. The $\nu_3$ frequency is probably the first overtone of $\nu_2$, but there does not seem to exist any aliasing or other logical link between $\nu_1$ and any of $\nu_2$ or $\nu_3$. It is remarkable here that $\nu_2$ is fully compatible with the suspected rotation frequency of $\lambda$~Cep near 0.25\,day$^{-1}$ (see Henrichs \& Sudnik \cite{HenSud2}), although, given the unstable behaviour of the variability of $\lambda$~Cep, this could also be a coincidence. An interesting point is that the $\nu_3$ frequency is quite close to the $0.46$\,day$^{-1}$ frequency seen in the power spectra of the absorption lines in the blue part of the spectrum (see above). This suggests that part of the variability of H$\alpha$ takes its origin in the same region as the absorption lines.
\begin{figure}
\resizebox{8cm}{!}{\includegraphics{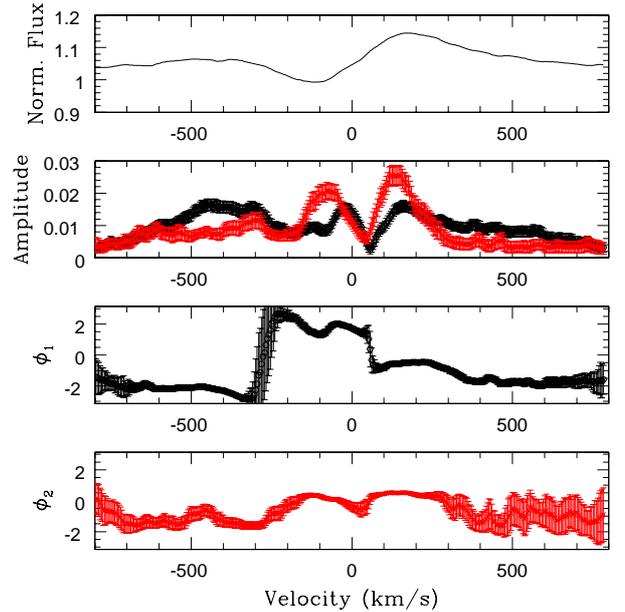}}
\caption{Variations of the H$\alpha$ emission line with the $\nu_2$ and $\nu_3$ frequencies in August 2013. From top to bottom, the panels illustrate the mean line profile, the amplitudes (black and red symbols stand for the variations at frequency $\nu_2$ and $\nu_3$, respectively), and the phase constants of the modulation as a function of wavelength.\label{nuHa}}
\end{figure}

\begin{figure}
\resizebox{8cm}{!}{\includegraphics{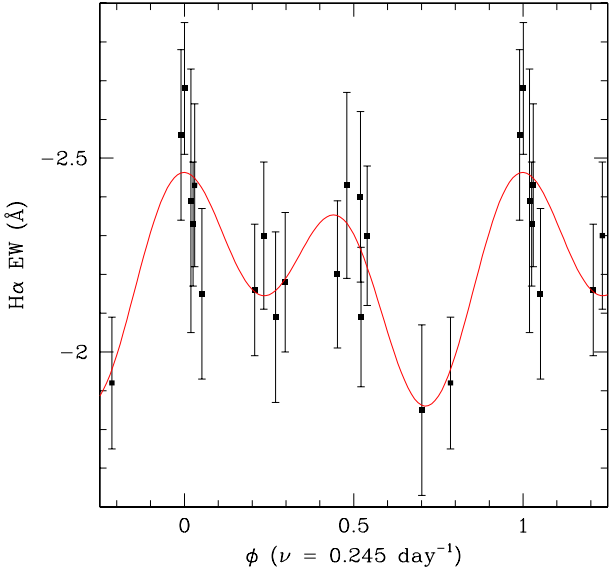}}
\caption{Variations of the EW of the H$\alpha$ line. Phase 0.0 corresponds to the time of the first optical spectrum of the August 2013 campaign (HJD~2\,456\,507.7607). The solid curve shows the expected variations as reconstructed from the amplitudes and phases of the $\nu_2$ and $2 \times \nu_2$ frequencies (see Fig.\,\ref{nuHa}), whilst the data points with error bars correspond to the actual EWs measured in August 2013.\label{EWha}}
\end{figure}

Assuming that $\nu_3$ is indeed the $n = 2$ harmonic frequency of $\nu_2$, we have used the information on phase constant and amplitude as a function of wavelength (Fig.\,\ref{nuHa}) to reconstruct the variations of the H$\alpha$ line within the cycle corresponding to $\nu_2$. We have then computed the equivalent width (EW) of the reconstructed profiles as a function of phase. The comparison with the observed EWs is shown in Fig.\,\ref{EWha}. For the observations, the errors were evaluated using the formalism of Vollmann \& Eversberg (\cite{VE}). Although the dispersion is quite large, the dominant trends of the data are indeed predicted by the results of the reconstruction.

Due to poor weather conditions at the TIGRE site, the June 2013 dataset is less densely sampled. The OHP data do not cover the H$\alpha$ line and we thus restrict the analysis of the June 2013 data to the \ion{He}{ii} $\lambda$\,4686 line. Because of the poorer sampling, the Fourier analysis is subject to a much more severe aliasing. The highest peak at $(1.405 \pm 0.02)$\,day$^{-1}$ accounts again for most of the power. Although it is close to $\nu_1$, the severe aliasing leads to a strong ambiguity about the actual frequency of the signal. 

\begin{figure}
\resizebox{8cm}{!}{\includegraphics{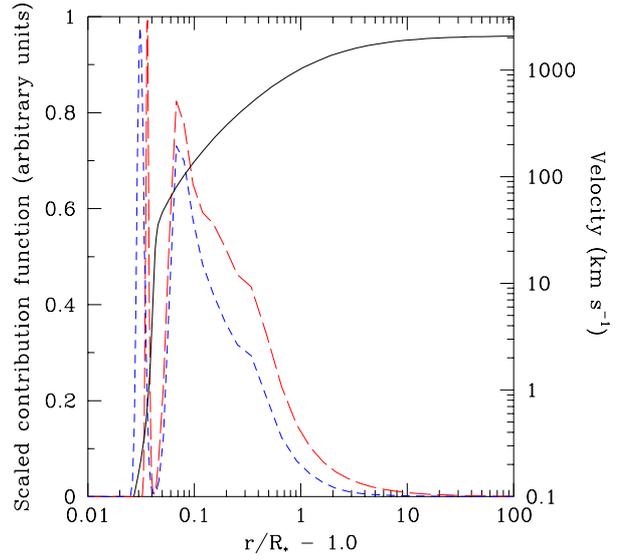}}
\caption{Normalized contribution functions showing the location of the line formation regions of the \ion{He}{ii} $\lambda$\,4686 (short-dashed line) and H$\alpha$ (long-dashed line) transitions according to the best-fit CMFGEN model of Bouret et al.\ (\cite{Bouret}). The solid line provides the corresponding velocity law.\label{spatial2}}
\end{figure}

Our finding of a very different behaviour of the \ion{He}{ii} $\lambda$\,4686 and H$\alpha$ lines of $\lambda$~Cep is surprising at first sight, as it would suggest that these emission lines arise from different parts of the wind. Under local thermodynamical equilibrium (LTE) conditions, one would expect the \ion{He}{ii} $\lambda$\,4686 to form closer to the photosphere than H$\alpha$ given the very different excitation potentials of the upper levels of the \ion{He}{ii} and the H$\alpha$ transitions (50.8 versus 12.1\,eV). However, the atmospheres of massive stars strongly deviate from LTE. For instance, the populations of the energy levels involved in the \ion{He}{ii} $\lambda$\,4686 transition are affected by pumping due to the \ion{He}{ii} Lyman lines below 300\,\AA. Figure\,\ref{spatial2} illustrates the contribution functions $\xi$ of both lines as a function of height above the photosphere as derived from the best-fit non-LTE CMFGEN model of Bouret et al.\ (\cite{Bouret}). The contribution function $\xi$ is defined by $$EW = \int_{R_*}^{\infty} \xi\,d\log{r}$$ (see Hillier \cite{Hillier89}). For the purpose of the discussion here, it is sufficient to consider the relative spatial variations of $\xi$. In Fig.\ \ref{spatial2}, we have thus normalized the contribution functions of both lines by dividing by their maximum value. In this spherically symmetric model, both lines actually form over similar regions, although the photospheric contribution to the \ion{He}{ii} $\lambda$\,4686 line arises deeper in the photosphere and this line's contribution function drops slightly faster in the wind than the one of H$\alpha$. Whether or not this slight difference is sufficient to explain the differences in the observed behaviour is unclear. In this context, Martins et al.\ (\cite{Martins2015}) recently studied the radial dependence of variability in seven O9 -- B0.5 stars. They found that lines that form in the photosphere can show very different variability patterns despite their line formation regions being close to each other. Our study indicates that this conclusion also applies to the case of the much earlier O6\,Ief star $\lambda$~Cep. 

In addition to the radial stratification discussed above, one might wonder whether the fact that $\lambda$~Cep is a fast rotator could lead to a latitudinal separation of the line formation regions. Indeed, in fast rotators such as HD~93\,521, gravity darkening is sufficient to produce a latitudinal separation of the formation regions of low-excitation absorption lines, which form near the cooler equator, and high-excitation lines, which are preferentially formed near the hotter polar regions (Rauw et al.\ \cite{RMP} and references therein). Adopting the Roche model, a rotational period of 4.1\,days and the stellar parameters (R$_*$, M$_*$) from Bouret et al.\ (\cite{Bouret}), we estimate that for $\lambda$~Cep the effective temperatures near the pole and near the equator would differ by 3\%, i.e.\ by about 1200\,K. This effect might also contribute to the observed difference in variability behaviour of the \ion{He}{ii} $\lambda$\,4686 and H$\alpha$ lines, although a decrease in temperature of about 5000\,K would be required to reach the temperature regime where the \ion{He}{ii} line is most sensitive to temperature.

\section{X-ray lightcurves \label{varX}}
Since X-ray emission from single non-magnetic O-type stars is commonly considered to arise from a distribution of shocks inside the stellar wind, one expects this X-ray emission to be variable on timescales of hours on which these shocks evolve. However, a comprehensive search for such short-term variability in the case of $\zeta$~Pup provided no significant detection (Naz\'e et al.\ \cite{zetaPup2}). This result implies a lower limit on the number of X-ray emitting parcels in the wind of $\zeta$~Pup of $10^5$ (Naz\'e et al.\ \cite{zetaPup2}). Still, Naz\'e et al.\ (\cite{zetaPup2}) found significant trends in the X-ray emission level on longer timescales of at least several days. These variations could be related to large-scale structures extending over a significant fraction of the wind and corotating with the star. In this respect, it is interesting to look for correlations between the optical and X-ray variability. Such a correlation between the H$\alpha$ emission strength and the {\it ROSAT} X-ray count rate in a narrow energy band had previously been reported by Bergh\"ofer et al.\ (\cite{BS}) for $\zeta$~Pup. The data from both wavelength ranges revealed a period of 16.7\,hrs. However, subsequent studies with {\it ASCA} (Oskinova et al.\ \cite{zetaOph}) and {\it XMM-Newton} (Kahn et al.\ \cite{Kahn}, Naz\'e et al.\ \cite{zetaPup2}) did not confirm the presence of this periodicity, suggesting that it was probably a transient feature. Another important result in this context was the recent discovery by Oskinova et al.\ (\cite{xiCMa}) of X-ray pulsations in the $\beta$~Cep-like pulsator $\xi^1$~CMa. The X-ray pulsations occur on the same period as the optical pulsations suggesting a direct connection between the two phenomena. Since both pulsations and large-scale wind structures have been reported for $\lambda$~Cep, we decided to take advantage of our data to study the variations of its X-ray flux. 

\begin{figure}
\resizebox{8cm}{!}{\includegraphics{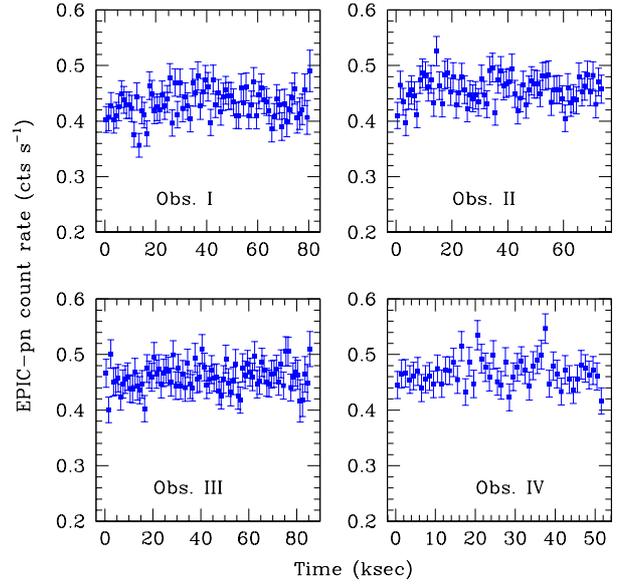}}
\caption{Background-corrected EPIC-pn X-ray lightcurves of $\lambda$~Cep with a time-bin of 1\,ks. The abscissae show the time elapsed since the start of each observation. \label{lcX}}
\end{figure}

We started by investigating the intra-pointing variability (see Fig.\,\ref{lcX}). Following the approach of Naz\'e et al.\ (\cite{zetaPup2}), we performed $\chi^2$ tests of the EPIC-MOS and pn background- and barycentric-corrected lightcurves of each observation. We did so for time-bins of 100\,s, 500\,s and 1\,ks, keeping only bins with more than 80\% fractional exposure time, and for several hypotheses (constancy, linear trend, quadratic trend). No significant intra-pointing variability is detected with these tests for observations I, II and IV. Significant, i.e.\ with significance level (SL) $<1$\%, intra-pointing variability is found for the pn data and bins of 100\,s in observation III. The MOS data of this observation have larger SL, hence show only moderate variability. However, for the 100\,s bins, the relative dispersions of the count rates about the mean are only marginally larger than what is expected from Poissonian photon noise. Some datasets apparently exhibit oscillations of the count rate. This is the case for instance of the pn data of obs.\ II (see Fig.\,\ref{lcX}) where a Fourier analysis yields a frequency of 4.48\,d$^{-1}$. However, this signal is not present in the MOS data and does not exist in the pn data of the other observations. 

\begin{table}[h]
\caption{EPIC count rates of $\lambda$~Cep in the 0.2 -- 10\,keV energy band averaged over the different observations.\label{CRs}}
\begin{center}
\begin{tabular}{c c c c}
\hline
Obs. & \multicolumn{3}{c}{Count rates (cts\,s$^{-1}$)} \\
\cline{2-4}
& MOS1 & MOS2 & pn \\
\hline
I   & $0.117 \pm 0.001$ & $0.119 \pm 0.001$ & $0.430 \pm 0.003$ \\
II  & $0.124 \pm 0.001$ & $0.124 \pm 0.001$ & $0.455 \pm 0.003$ \\
III & $0.125 \pm 0.001$ & $0.126 \pm 0.001$ & $0.459 \pm 0.003$ \\
IV  & $0.126 \pm 0.002$ & $0.129 \pm 0.002$ & $0.466 \pm 0.003$ \\
\hline
\end{tabular}
\end{center}
\tablefoot{The 1$\sigma$ uncertainties of the mean are quoted.}
\end{table}

\begin{figure}
\resizebox{8cm}{!}{\includegraphics{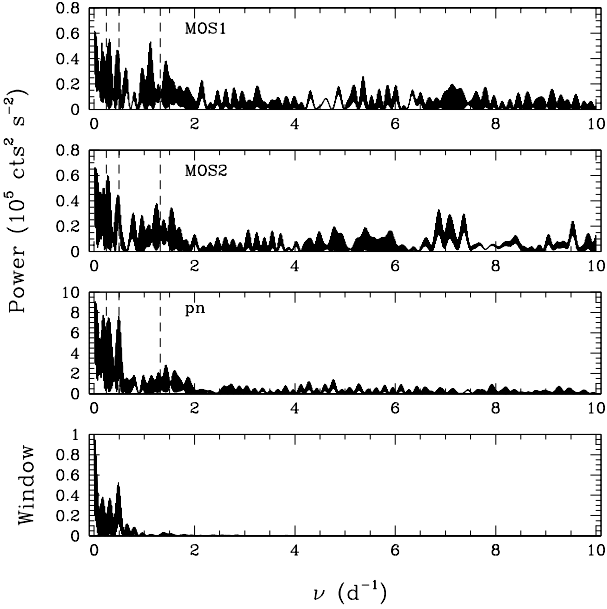}}
\caption{Fourier analysis of the X-ray lightcurves of $\lambda$~Cep in June and August 2013. The top three panels show the power spectra for the three EPIC instruments whilst the bottom panel displays the power spectral window of the EPIC-pn time series. MOS1 and MOS2 have very similar spectral windows. The dashed vertical lines indicate the frequencies $\nu_1 = 1.315$, $\nu_2 = 0.245$ and $\nu_3 = 0.495$\,d$^{-1}$ found in the analysis of our optical spectra.\label{FourierX}}
\end{figure}

Since the individual pointings last typically for a bit less than one day, the above tests will only reveal variability on timescales shorter than one day. As a next step we have then performed a Fourier analysis of the full (i.e.\ all four observations) lightcurves with 1\,ks bins. For this purpose we used again the method of Heck et al.\ (\cite{HMM}) amended by Gosset et al.\ (\cite{Gosset}). To get the cleanest results, we again discarded time bins with less than 80\% fractional exposure time. Fig.\,\ref{FourierX} illustrates the results of this Fourier analysis for all four observations. The highest peaks are found at frequencies below 0.5\,d$^{-1}$. However, the positions of the highest peaks and the general shape of the periodograms, especially above 2\,d$^{-1}$, are quite different for the three EPIC instruments. As pointed out by Naz\'e et al.\ (\cite{zetaPup2}), having three independent cameras onboard {\it XMM-Newton} that operate simultaneously is extremely beneficial to check the consistency of a timing analysis. The differences in Fig.\,\ref{FourierX}, especially at high frequencies, most probably reflect different realizations of the noise, whilst the low-frequency content of the power spectra hints at the existence of variations with timescales larger than two days. 

\begin{figure}
\resizebox{8cm}{!}{\includegraphics{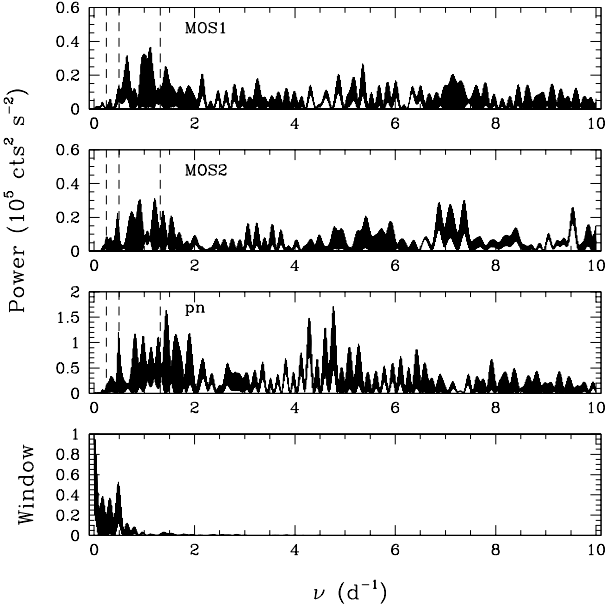}}
\caption{Same as Fig.\,\ref{FourierX}, but after subtracting the long-term variations.\label{FourierXcorr}}
\end{figure}

The existence of a long- or medium-term variability can also be seen from Table\,\ref{CRs} which shows that the mean EPIC count rates of $\lambda$~Cep are about 7\% lower in observation I than in later observations, while a shallower (about 2 -- 3\%) increase occurs from observation II to IV. The same trends are also seen in the fluxes derived from the global spectral fits in Table\,\ref{xspecfits}. To check whether or not these trends affect the detection of shorter periodicities, we have subtracted the average count rate from the lightcurve of individual observations and repeated our Fourier analysis. The corresponding periodogram is displayed in Fig.\,\ref{FourierXcorr}. The situation is qualitatively similar to that of Fig.\,\ref{FourierX}, except of course for the lack of power at frequencies below about 0.5\,d$^{-1}$. In conclusion, our X-ray dataset alone cannot constrain the timescale of the long-term variations and no unambiguous short-term periodicity can be identified.

\begin{figure}
\resizebox{8cm}{!}{\includegraphics{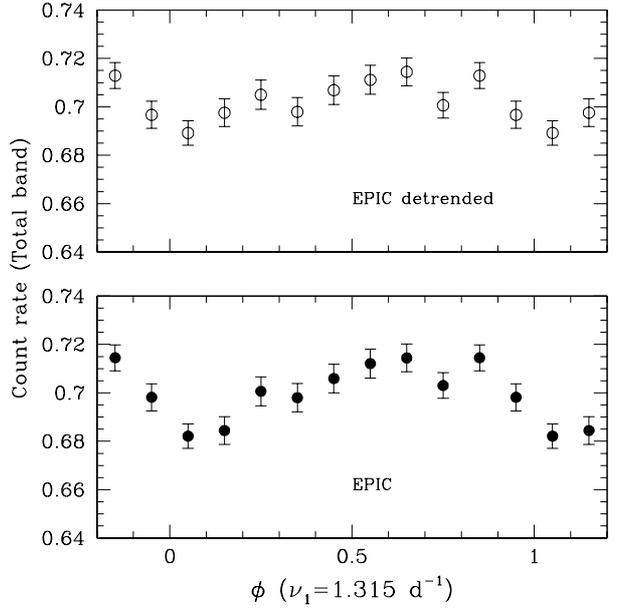}}
\caption{X-ray lightcurve of the full EPIC (MOS1 + MOS2 + pn) count rate folded in phase with the $\nu_1$ frequency. The data were binned into phase intervals of 0.1 and phase 0.0 was arbitrarily taken as the starting time of observation I (HJD~2\,456\,456.1871). The bottom and top panels show the results for the background-corrected count rates and after removing the long-term trend from the background-corrected count rates, respectively (see text).\label{foldXnu1}}
\end{figure}

We can nevertheless fold the X-ray lightcurve with the frequencies detected in our analysis of the simultaneous optical spectroscopy. Starting with the $\nu_1 = 1.315$\,d$^{-1}$ frequency found in the variations of our \ion{He}{ii} $\lambda$\,4686 data, we note a group of peaks between 1.1 and 1.4\,day$^{-1}$ in Fig.\,\ref{FourierXcorr}. These peaks are probably not significant as their frequencies differ between the three EPIC instruments. Figure\,\ref{foldXnu1} displays the X-ray lightcurve for both the observed and the detrended\footnote{Since the 18.3\,hrs period corresponding to $\nu_1$ is shorter than the duration of an individual observation, we consider also the results after taking out the long-term variations.} count rates phased with $\nu_1$. The peak-to-peak amplitude of the modulation is 3 -- 5 \% of the EPIC count rate whilst the (one-sided) error bars are 0.8\%. The amplitude of any modulation of the X-ray flux with $\nu_1$ is thus significant at the 2 -- 3\,$\sigma$ level only, in agreement with the results obtained in our Fourier analysis.         

\begin{figure}
\resizebox{8cm}{!}{\includegraphics{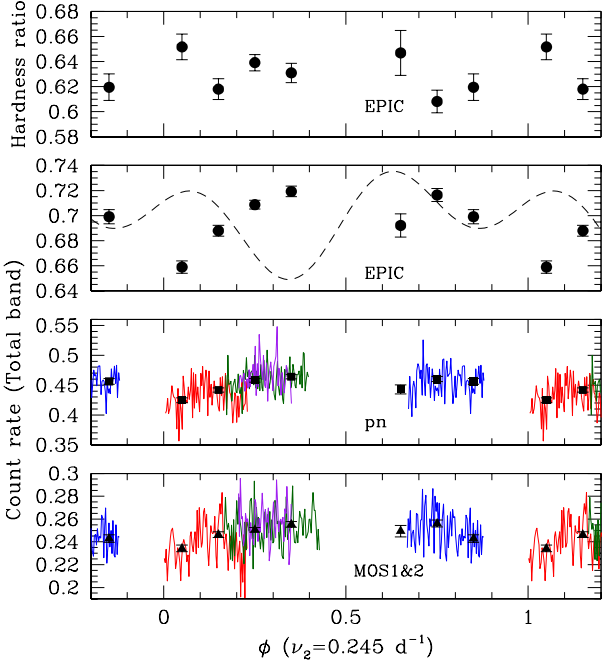}}
\caption{X-ray lightcurve of $\lambda$~Cep folded in phase with the $\nu_2$ frequency. The red, blue, green and purple curves correspond to the background-corrected count rates as measured during observations I, II, III and IV respectively. These lightcurves were binned into 1\,ks bins. The filled symbols with error bars indicate the lightcurve binned into phase intervals of 0.1. Phase 0.0 was again taken as the starting time of observation I. From bottom to top the results are shown for the combination of the two MOS detectors, for the pn, for the combination of all three EPIC instruments and for the hardness ratio derived for all three EPIC cameras. The dashed line in the second panel from the top represents the variations of the EW of the H$\alpha$ line scaled to match the range of variability of the EPIC count rate and shifted by 0.36 in phase (with respect to Fig.\,\ref{EWha}) to match the phase 0.0 of the folded X-ray lightcurve.\label{foldXnu2}}
\end{figure}

We have then considered the $\nu_2 = 0.245$\,d$^{-1}$ frequency which we found along with its first overtone in the variations of the H$\alpha$ line profile. Figure\,\ref{foldXnu2} displays the corresponding lightcurve built from the background-corrected count rates. For the EPIC count rates, the peak-to-peak amplitude now amounts to $\sim 9$\% for a mean 0.8\% one-sided error bar. This results in a signal significant at the 5\,$\sigma$ level. This modulation mainly stems from the long-term trend discussed above. We have also folded the hardness ratio, computed as the ratio of the EPIC count rates in the hard band over those in the soft band, with the $\nu_2$ frequency. Whilst the source appears harder when the X-ray emission is minimium, this result is not significant in view of the large error bars.

Our data sample only 65\% of the full 4.1\,day cycle, and additional data are clearly required to establish this periodicity. Still, we note that the shape of the EPIC lightcurve in the second panel of Fig.\,\ref{foldXnu2} suggests that the modulation could have a double-wave shape. Figure\,\ref{foldXnu2} further shows the predicted behaviour of the EW of the H$\alpha$ line. The comparison indicates that the highest X-ray flux is observed during the secondary minimum of the EW of the optical emission line. This would suggest a close connection, although shifted in phase, with the mechanism responsible for the H$\alpha$ variability. A similar situation was found by Bergh\"ofer et al.\ (\cite{BS}) for the H$\alpha$ and soft X-ray emission of $\zeta$~Pup, although more recent X-ray observations did not confirm the existence of the 16.7\,hrs period. Bergh\"ofer et al.\ (\cite{BS}) argued that the phase shift between the two energy domains was consistent with the wind flow time between the H$\alpha$ and X-ray emission regions. In the present case, this scenario does not seem to work as the X-ray and H$\alpha$ formation regions overlap to a large extent (see the forthcoming Sect.\,\ref{global}) and wind flow times would be much shorter than the observed time lag. Instead, the phase shift could rather suggest an anti-correlation between the X-ray and H$\alpha$ emission, e.g.\ due to an extra heating mechanism.   

In summary, we thus conclude that the X-ray flux of $\lambda$~Cep does not undergo highly significant variations on timescales of a few hours, but could undergo a rotational modulation on timescales of about 4.1\,days. This result however calls for confirmation through additional X-ray observations.

$\lambda$~Cep is not the only single O-type star for which a rotational modulation of the X-ray emission has possibly been detected. Indeed, in addition to the case of $\zeta$~Pup (Naz\'e et al.\ \cite{zetaPup2}) discussed above, such a modulation has been claimed in the case of $\zeta$~Oph (O9.5\,V) and of $\xi$~Per (O7.5\,III(n)((f))). In the former case, Oskinova et al.\ (\cite{zetaOph}) found that the X-ray flux measured with {\it ASCA} over 1.185\,days displays a 20\% modulation on a period of 0.77\,days (i.e.\ about half the probable rotational period). In the latter case, {\it Chandra}-HETG data cover a bit less than half the probable rotational period of 4.2\,days and they display a 15\% variation. Massa et al.\ (\cite{Massa}) suggest that the putative rotational modulation of $\xi$~Per's X-ray flux stems from so-called corotating interaction regions (CIRs). These authors note the absence of hardness variations of the X-ray emission of $\xi$~Per and accordingly conclude that the variations of the X-ray flux stem from obscuration by optically thick structures.  

\section{The X-ray spectrum of $\lambda$~Cep \label{sectX}}
In this section, we analyse the X-ray spectrum of $\lambda$~Cep using techniques of increasing complexity and sophistication. Whenever appropriate, we compare our results with those obtained by Bouret et al.\ (\cite{Bouret}) from a combined far-UV, UV and optical spectral analysis of $\lambda$~Cep using the CMFGEN model atmosphere code (Hillier \& Miller \cite{CMFGEN}) accounting for the effect of micro-clumping. Bouret et al.\ (\cite{Bouret}) derived an effective temperature of 36\,000\,K, a gravity of $\log{g} = 3.5$, a mass-loss rate of $\log{\dot{M}} = -5.85$ and a wind velocity of 2100\,km\,s$^{-1}$. They further determined a moderate chemical enrichment with $y/y_{\odot} = 1.33$, $\epsilon(N)/\epsilon(N)_{\odot} = 8.3 \pm 3.0$. Carbon and oxygen were found to be slightly depleted: $\epsilon(C)/\epsilon(C)_{\odot} = 0.68 \pm 0.34$, $\epsilon(O)/\epsilon(O)_{\odot} = 0.66 \pm 0.23$, although, within the error bars, these numbers are essentially consistent with the corresponding solar abundances of Asplund et al.\ (\cite{Sun3D}). 

\begin{figure*}[htb]
\resizebox{15cm}{!}{\includegraphics{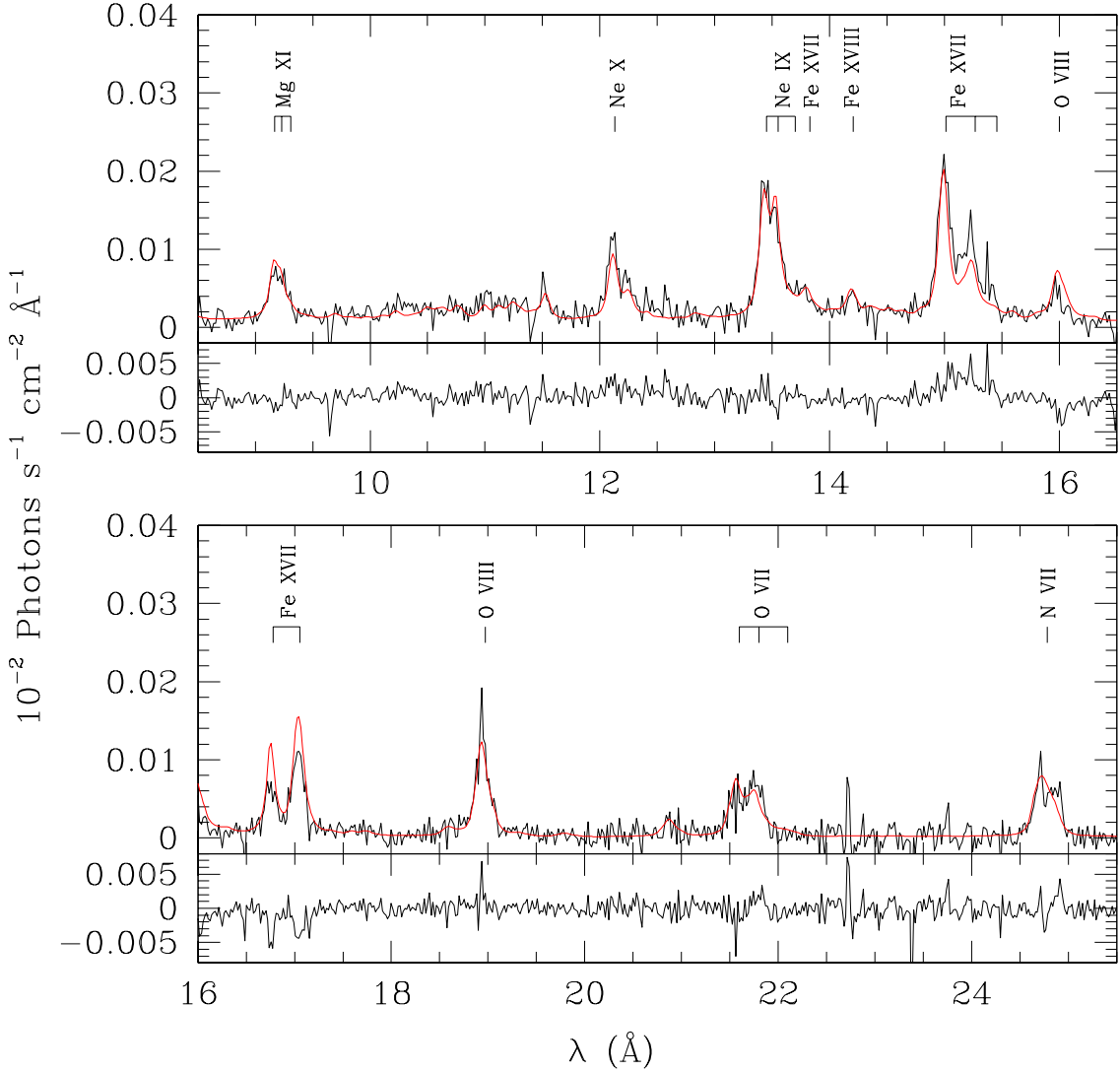}}
\caption{The RGS spectrum of $\lambda$~Cep. The most prominent lines are identified with the labels. The red line shows our best fit with the model of Sect.\,\ref{global}. The lower panels show the differences between the data and the model.\label{RGSlCep}}
\end{figure*}

\subsection{The RGS spectrum \label{RGSspec}}
The RGS spectrum (Fig.\,\ref{RGSlCep}) provides exploitable data over the wavelength range between 9 and 25\,\AA. At the shorter wavelength end, the flux intrinsically drops off because of the moderate plasma temperatures, leading to too low a S/N ratio in the RGS spectrum. Conversely, at wavelengths above 25\,\AA, the X-ray emission of $\lambda$~Cep is heavily absorbed by the interstellar medium and, to a lesser extent, by the cool material in the stellar wind.

Although rather noisy, the RGS spectrum reveals a number of emission lines that can be used as diagnostics of the hot plasma in the wind of $\lambda$~Cep. Beside several strong \ion{Fe}{xvii} lines, the strongest features are the Ly$\alpha$ lines of \ion{N}{vii}, \ion{O}{viii} and \ion{Ne}{x}, and the He-like triplets of \ion{O}{vii}, \ion{Ne}{ix} and \ion{Mg}{xi}. 

The ratio between the fluxes of the hydrogen and helium-like ions of oxygen and neon can be used to constrain the temperature of the emitting plasma. To do so, we need to correct the fluxes for the wavelength dependence of the interstellar and wind absorption. For this purpose, we adopt the interstellar column density of $0.25 \times 10^{22}$\,cm$^{-2}$ determined by Bohlin et al.\ (\cite{Bohlin}) and we use the ionized wind column (Naz\'e et al.\ \cite{HD108}) of $\log{N_{\rm wind}} = 21.79$ as derived from the spectral fits in Sect.\,\ref{broadband}. The absorption-corrected flux ratios are $3.15^{+1.57}_{-1.06}$ and $0.48^{+0.24}_{-0.16}$ for the \ion{Ne}{ix} triplet / \ion{Ne}{x} Ly$\alpha$ and \ion{O}{vii} triplet / \ion{O}{viii} Ly$\alpha$ ratios, respectively. These numbers yield plasma temperatures of $kT = (0.33 \pm 0.03)$ and $(0.31 \pm 0.04)$\,keV for the neon and oxygen ions, thereby indicating that a significant part of the X-ray emission of $\lambda$~Cep arises from a plasma at a temperature near 0.3\,keV. 

The strength of the blend between \ion{N}{vii} Ly$\alpha$ and \ion{N}{vi} $\lambda$\,24.9 compared to the \ion{O}{viii} Ly$\alpha$ line further suggests an N/O ratio higher than solar. Comparing the emissivities of the nitrogen and oxygen lines with the absorption-corrected line strength ratio ($0.40^{+.20}_{-.13}$) as measured on our data, we estimate an N/O ratio that would be typically $5.0^{+1.5}_{-1.0}$ times solar if the nitrogen emission arises from plasma at the same temperature as the oxygen lines. This result is somewhat lower than the enrichment found by Bouret et al.\ (\cite{Bouret}), but confirms indeed the chemical enrichment of the wind of $\lambda$~Cep.   

The He-like triplets consist of a resonance line ($r$), an intercombination doublet ($i$) and a forbidden line ($f$). In the case of $\lambda$~Cep and other massive stars, the $f$ components are strongly suppressed, whereas the $i$ components are enhanced compared to what one would expect from the transition probabilities. This situation stems from the harsh photospheric UV radiation field (Gabriel \& Jordan \cite{GJ}, Blumenthal et al.\ \cite{Blumenthal}, Porquet et al.\ \cite{Porquet}, Leutenegger et al.\ \cite{Leutenegger06}) that pumps the electrons from the upper level of the $f$ transition (2\,$^3S_1$) to the upper level of the $i$ doublet (2\,$^3P_{1,2}$). For instance, the measured ${\cal R} = f/i$ ratios of the \ion{O}{vii} and \ion{Ne}{ix} triplets are $\leq 0.12$ and $0.30 \pm 0.18$, respectively whilst the values expected in the absence of UV pumping or collisional depopulation of the 2\,$^3S_1$ level would be about 3.8 and 3 (Porquet et al.\ \cite{Porquet}).     

\begin{table*}
\caption{Spectral parameters for a simultaneous fit of the EPIC and RGS spectra \label{xspecfits}}
\begin{center}
\begin{tabular}{c c c c c c c c c c}
\hline
Obs. & $\log{N_{\rm wind}}$ & $kT_1$ & norm$_1$ & $kT_2$ & norm$_2$ & $(N/H)/(N/H)_{\odot}$ & $\chi^2_{\nu}$ (d.o.f) & $f_X^{\rm obs}$ & $f_X^{\rm cor}$\\
\cline{9-10}
& (cm$^{-2}$) & (keV) & (cm$^{-5}$) & (keV) & (cm$^{-5}$) & & & \multicolumn{2}{c}{$10^{-13}$\,erg\,cm$^{-2}$\,s$^{-1}$}\\
\hline
\vspace*{-3mm}\\
I   & $21.78^{+.01}_{-.02}$ & $0.283^{+.004}_{-.004}$ & $(2.87^{+.20}_{-.21})\,10^{-3}$ & $0.86^{+.02}_{-.04}$ & $(4.12^{+.43}_{-.20})\,10^{-4}$ & $3.88^{+.60}_{-.56}$ & 1.67 (1253) & $6.90 \pm 0.10$ & 16.7 \\
\vspace*{-3mm}\\
II  & $21.81^{+.02}_{-.02}$ & $0.285^{+.003}_{-.003}$ & $(3.68^{+.29}_{-.38})\,10^{-3}$ & $0.86^{+.01}_{-.01}$ & $(3.75^{+.20}_{-.17})\,10^{-4}$ & $2.91^{+.54}_{-.37}$ & 1.78 (1265) & $7.24 \pm 0.09$ & 17.8 \\
\vspace*{-3mm}\\
III  & $21.79^{+.02}_{-.01}$ & $0.286^{+.003}_{-.003}$ & $(3.27^{+.31}_{-.20})\,10^{-3}$ & $0.86^{+.02}_{-.02}$ & $(4.07^{+.25}_{-.19})\,10^{-4}$ & $3.14^{+.47}_{-.44}$ & 1.76 (1370) & $7.33 \pm 0.10$ & 17.8 \\
\vspace*{-3mm}\\
IV  & $21.79^{+.02}_{-.02}$ & $0.289^{+.004}_{-.004}$ & $(3.31^{+.37}_{-.26})\,10^{-3}$ & $0.86^{+.01}_{-.02}$ & $(4.01^{+.28}_{-.25})\,10^{-4}$ & $3.41^{+.63}_{-.60}$ & 1.45 (1279) & $7.45 \pm 0.10$ & 18.0 \\
\vspace*{-3mm}\\
\hline
\end{tabular}
\end{center}
\tablefoot{The RGS data are fitted between 0.4 and 1.5\,keV. The expression of the {\tt xspec} model is {\tt wabs*mtab\{wind\}*apec(2T)}. The interstellar neutral hydrogen column density is fixed to $0.25 \times 10^{22}$\,cm$^{-2}$. The helium abundance by mass $y$ is set to 1.33 solar. Except for nitrogen, all other abundances are kept solar. The last two columns provide the observed (last-but-one column) and ISM corrected (last column) fluxes in the energy band 0.5 -- 10\,keV. The normalization of the models is given as $10^{-14}\,\frac{\int n_e\,n_H\,dV}{4\,\pi\,d^2}$ where $n_e$ and $n_H$ are the electron and proton densities of the X-ray emitting plasma in cm$^{-3}$, and $d$ is the distance in cm. The quoted error bars correspond to the 90\% confidence level.}
\end{table*}
\begin{figure*}[t!]
\begin{minipage}{4.5cm}
\resizebox{4.5cm}{!}{\includegraphics{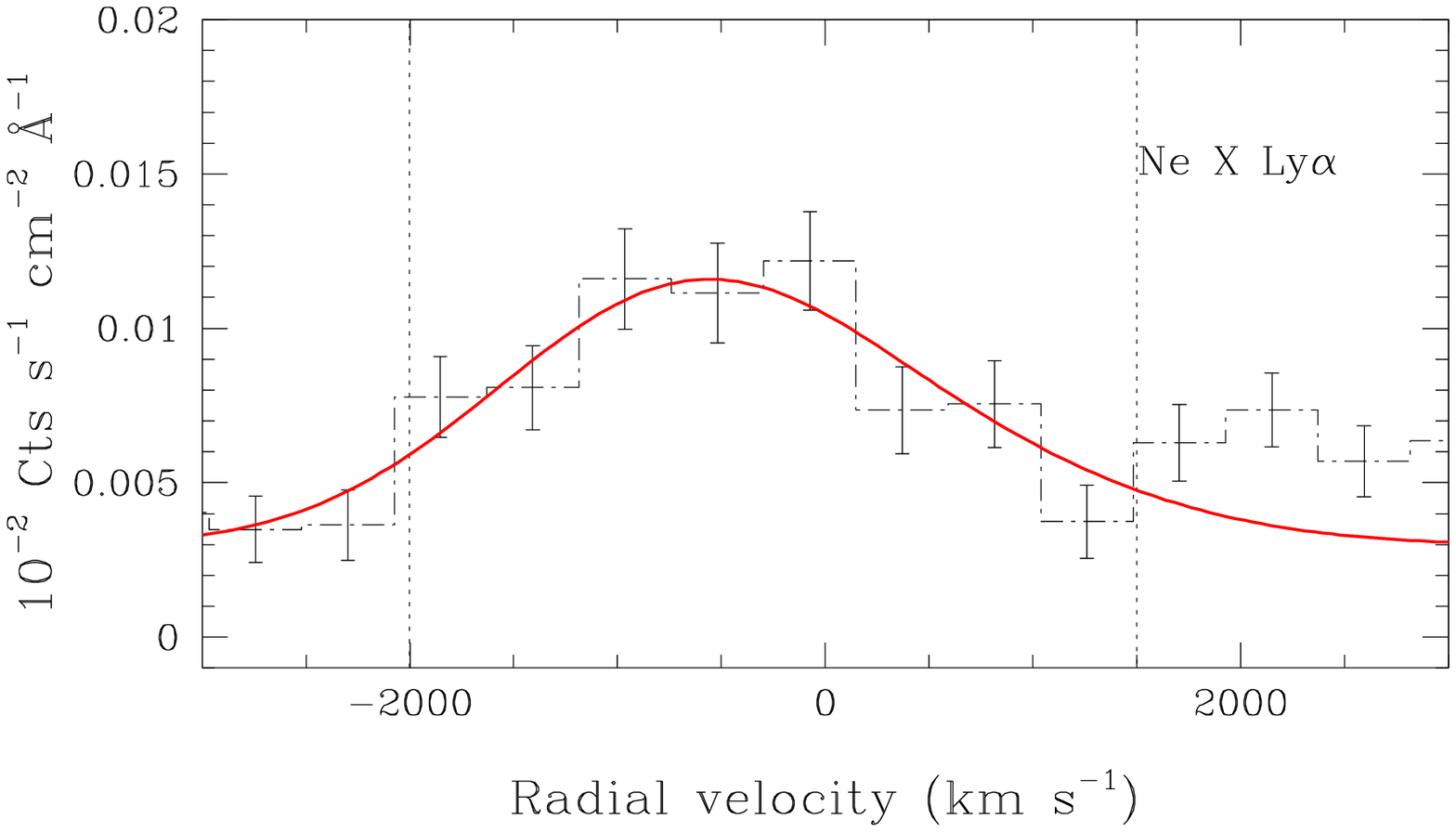}}
\resizebox{4.5cm}{!}{\includegraphics{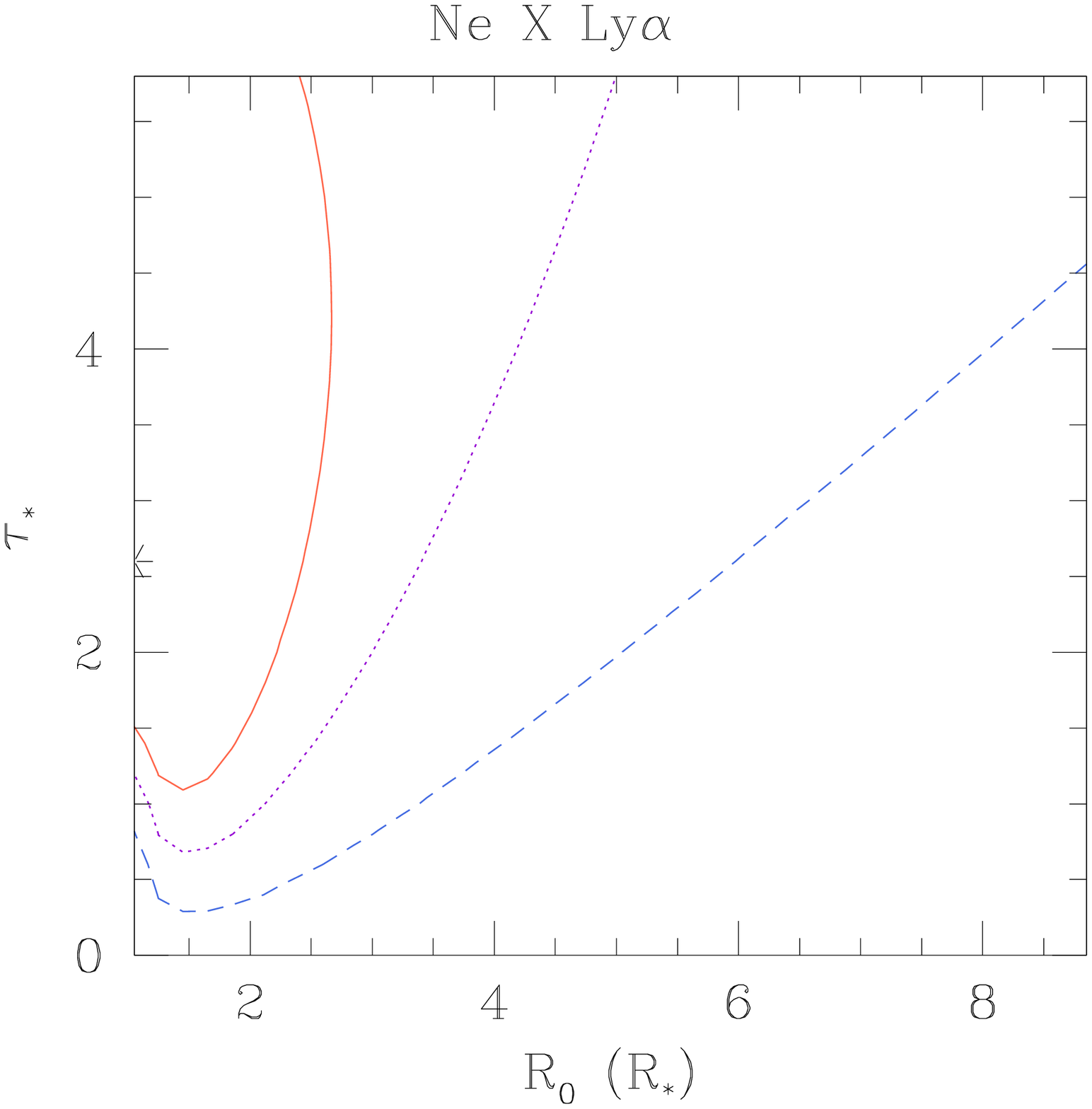}}
\end{minipage}
\begin{minipage}{4.5cm}
\resizebox{4.5cm}{!}{\includegraphics{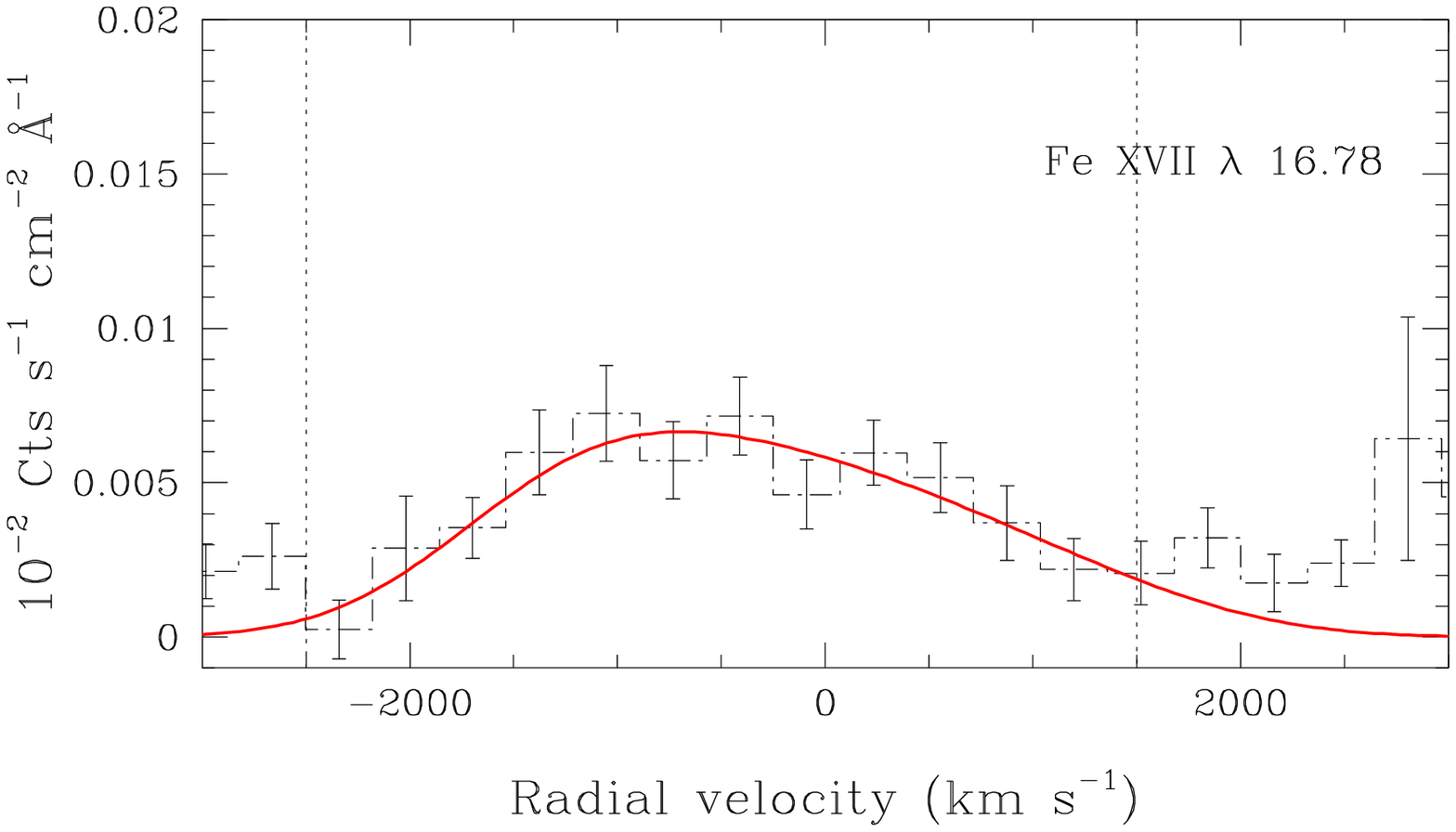}}
\resizebox{4.5cm}{!}{\includegraphics{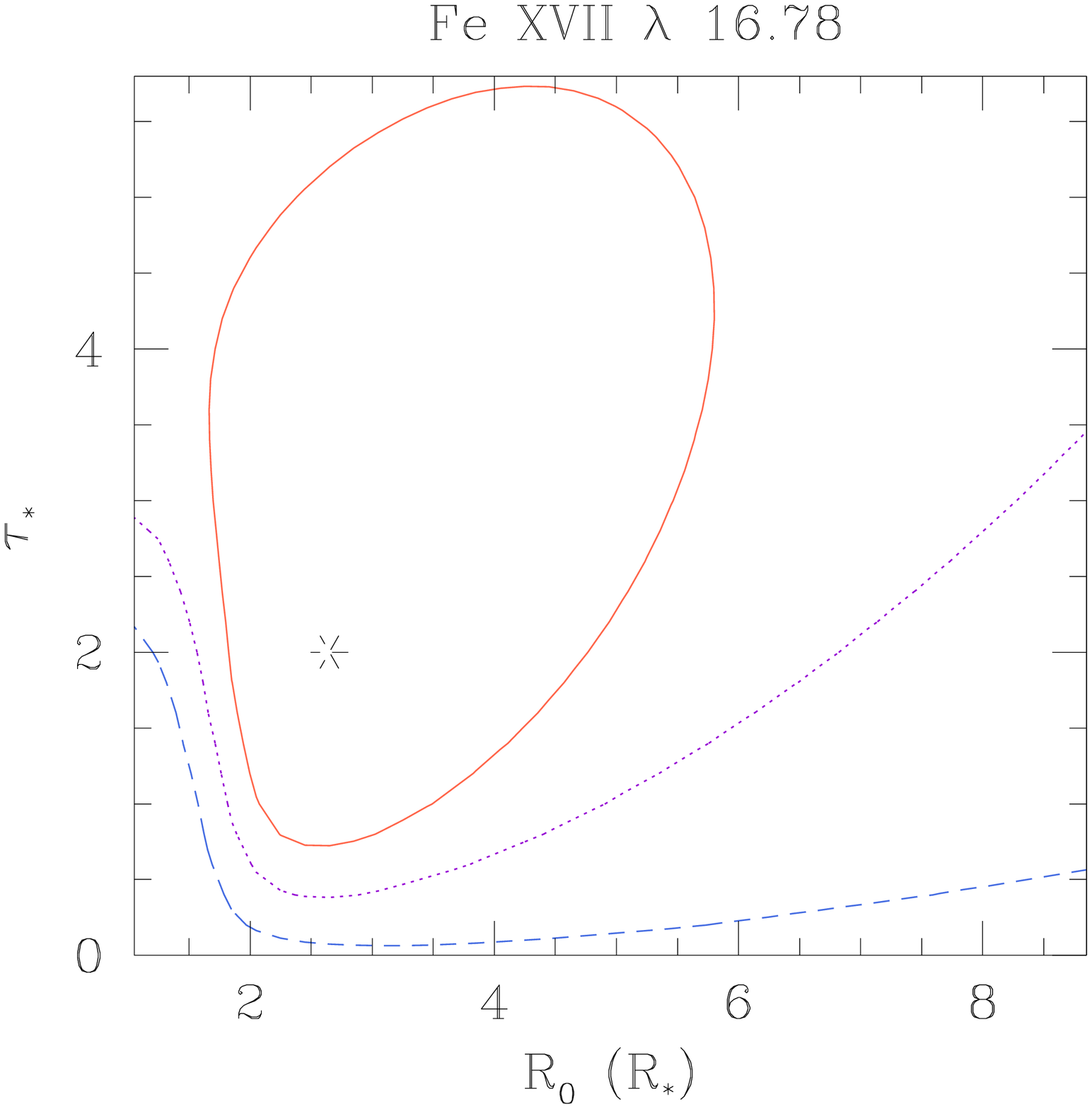}}
\end{minipage}
\begin{minipage}{4.5cm}
\resizebox{4.5cm}{!}{\includegraphics{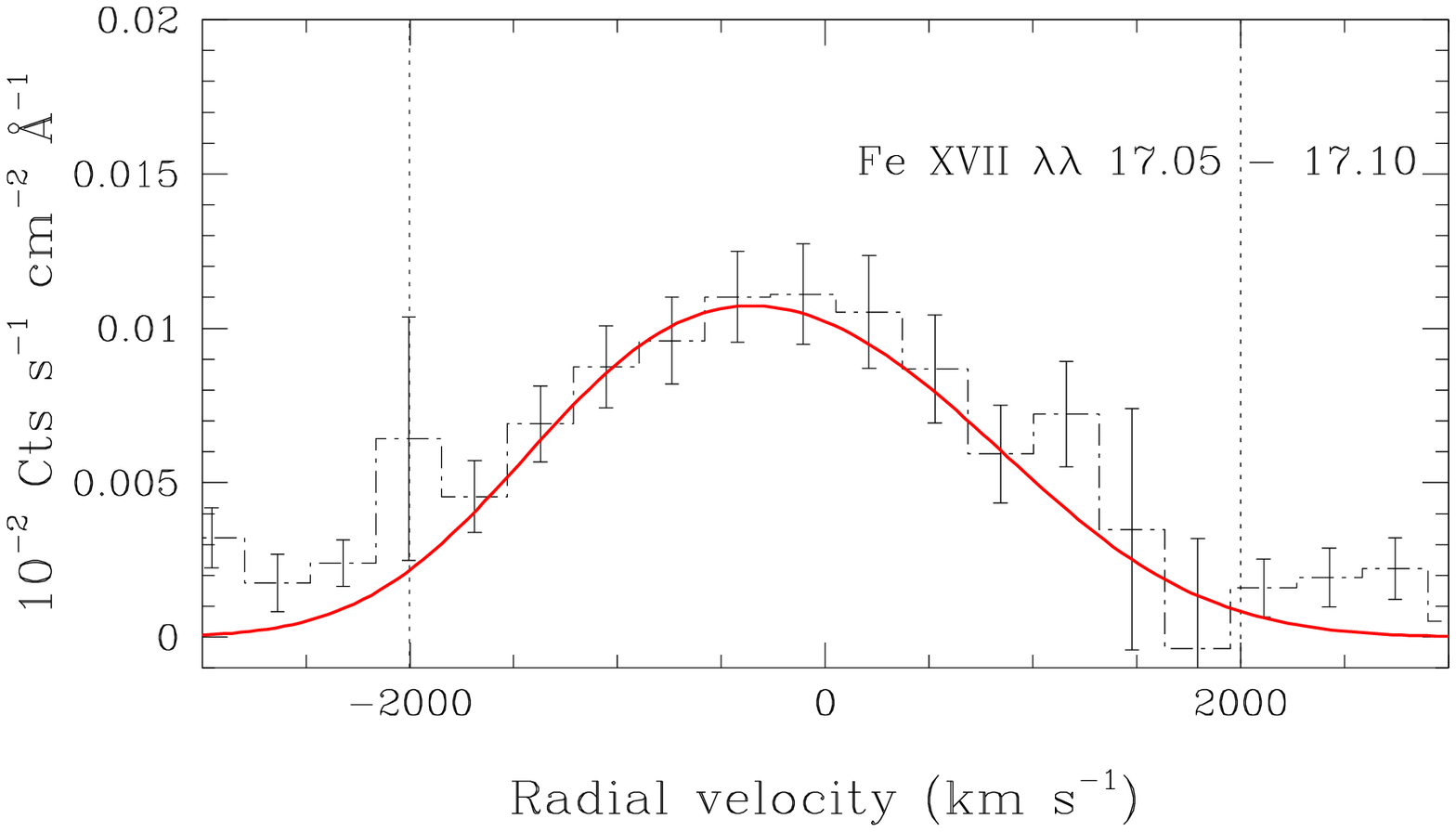}}
\resizebox{4.5cm}{!}{\includegraphics{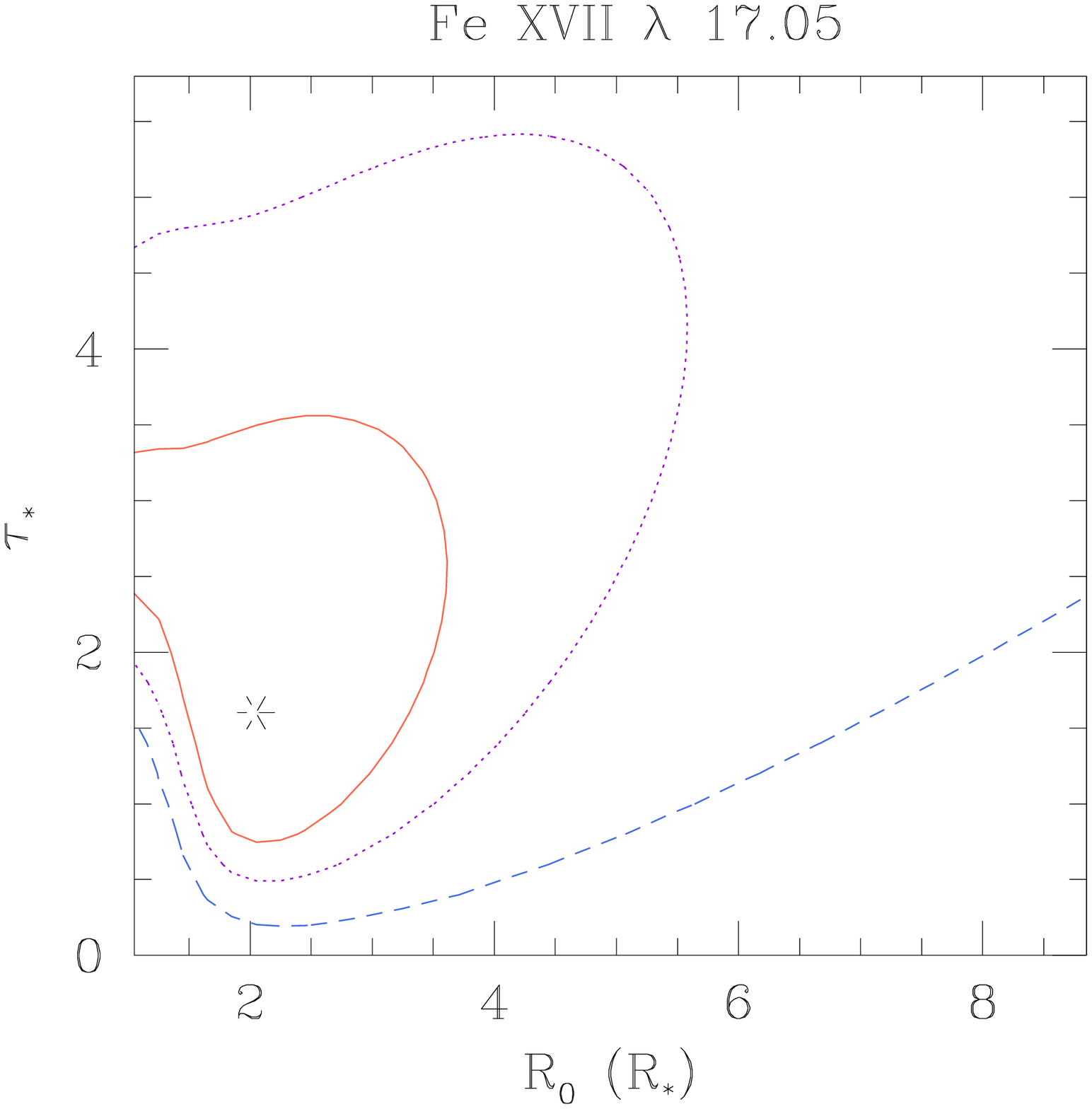}}
\end{minipage}
\begin{minipage}{4.5cm}
\resizebox{4.5cm}{!}{\includegraphics{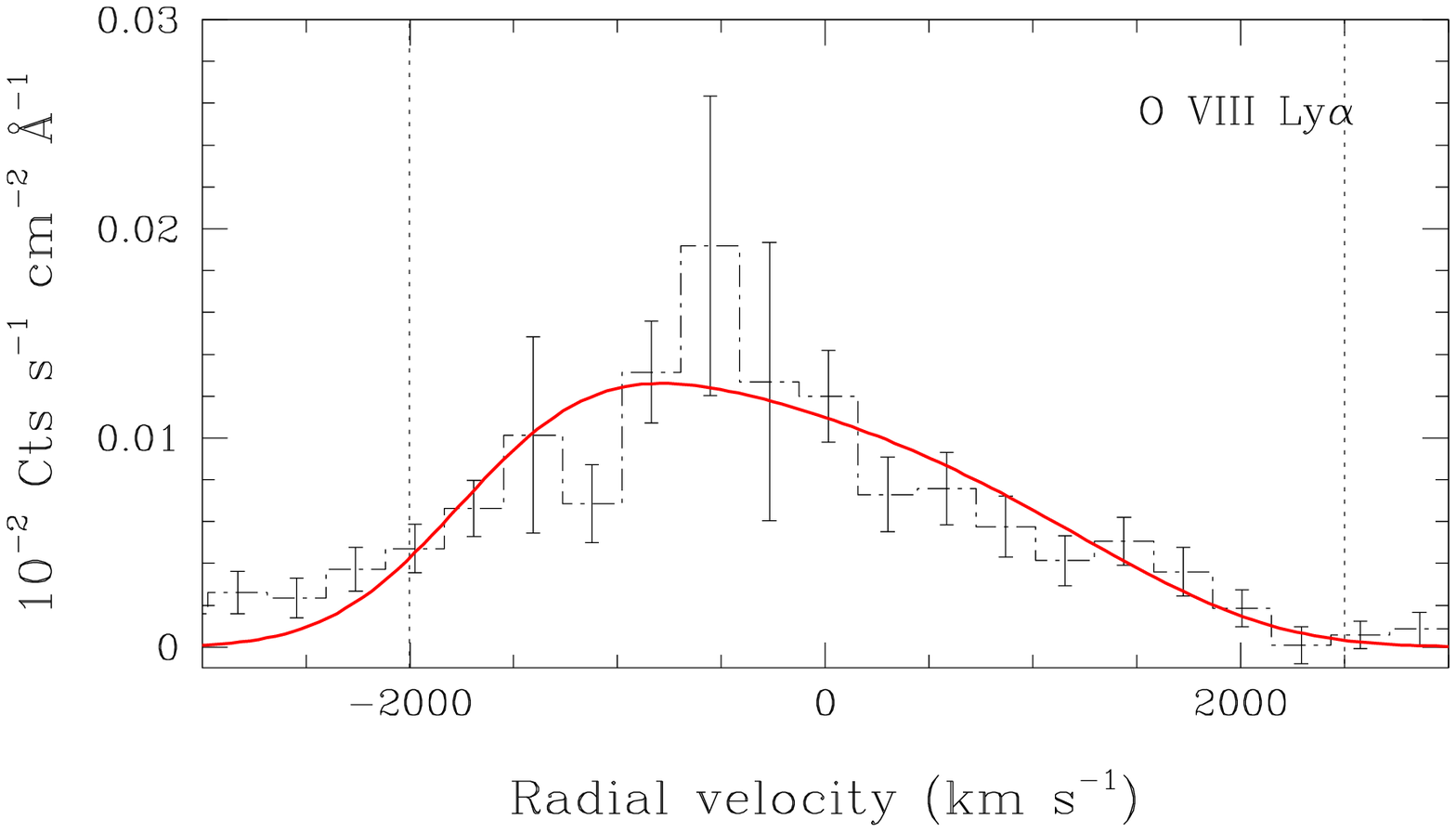}}
\resizebox{4.5cm}{!}{\includegraphics{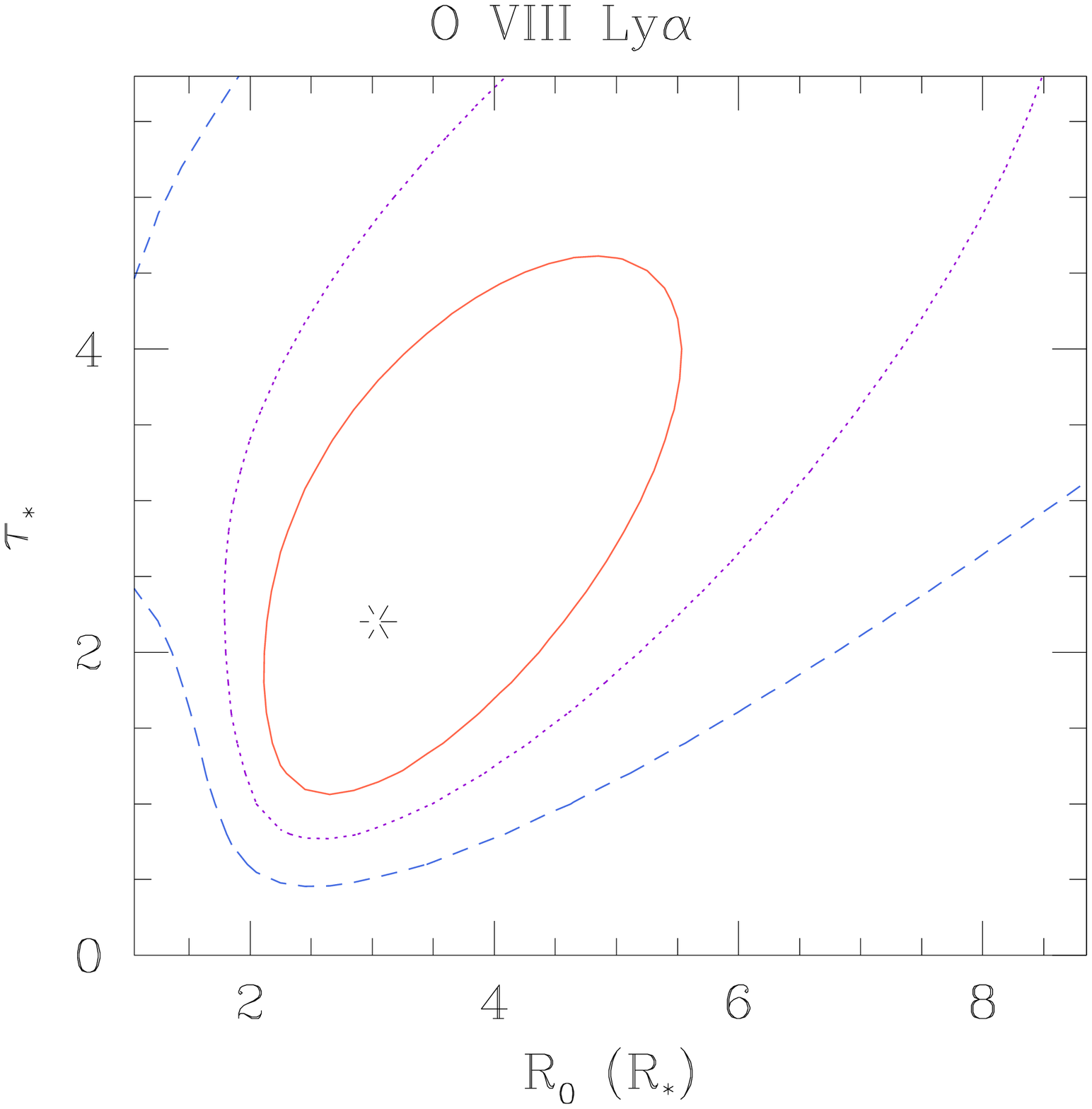}}
\end{minipage}
\caption{Fits of four individual lines in the RGS spectrum of $\lambda$~Cep. For each row, the top panel illustrates the best-fit model convolved with the RGS response and compared with the observed profile. The bottom panel shows the 68.3 (1$\sigma$, red continous line), 90 (violet dotted line) and 99\% (blue dashed line) confidence contours in the $\tau_*$ versus $R_0$ parameter space. The asterisks indicate the best-fit parameters.\label{fits}}
\end{figure*}

\subsection{Global fits of the X-ray spectrum \label{broadband}}
Using the {\tt xspec} software (Arnaud \cite{Arnaud}), we have attempted to fit the broadband X-ray spectra of $\lambda$~Cep with {\tt apec} optically-thin thermal plasma models (Smith \& Brickhouse \cite{apec}) absorbed by the interstellar medium (with the neutral hydrogen column density fixed to $0.25 \times 10^{22}$\,cm$^{-2}$) and by an ionized stellar wind using the model of Naz\'e et al.\ (\cite{HD108})\footnote{The stellar wind absorption was imported into {\tt xspec} as a multiplicative tabular model ({\tt mtable}).}. For this purpose, we have considered various combinations of the data: 
\begin{enumerate}
\item The EPIC spectra of each individual observation.
\item The EPIC and RGS spectra of each individual observation. The RGS spectra were considered over the energy range 0.4 -- 1.5\,keV.
\end{enumerate}
In each case, we tested three different assumptions on the plasma abundances: 
\begin{enumerate}
\item Solar abundances for all elements following Asplund et al.\ (\cite{Asplund}).
\item Abundances of He, C, N and O as derived by Bouret et al.\ (\cite{Bouret}), all other elements having solar abundances as above.
\item He abundance fixed to 1.33 times solar following Bouret et al.\ (\cite{Bouret}) and free N abundance, all other elements having solar abundances. 
\end{enumerate}

From these tests, we immediately found that the broadband spectrum of $\lambda$~Cep requires two plasma components to be represented by the model. The temperatures of these components are consistently found to be $kT_1 = 0.286 \pm 0.003$\,keV for the softer component and $kT_2 = 0.86 \pm 0.01$\,keV for the harder one. $kT_1$ is usually about 0.01\,keV lower in the solar abundance case, but apart from this small effect, the temperatures are rather independent of the assumed composition. They are also independent of what observation we consider (see Table\,\ref{xspecfits}). The best-fit wind column density $\log{N_{\rm wind}} = 21.79 \pm 0.01$ is also found to be very stable, both with respect to the assumed chemical composition and the observation under consideration. Finally, we found that the best fit quality was always achieved by the models where the He abundance by mass was set to 1.33 times solar and the N abundance was left as a free parameter. Its best-fit value was found to be $3.3 \pm 0.4$ solar. This is about half the value derived by Bouret et al.\ (\cite{Bouret}) but is consistent with the value derived from the strengths of individual lines (see Sect.\,\ref{RGSspec}). Adopting the CNO abundances as proposed by these authors resulted in a poorer fit of the X-ray spectrum. In this context, we stress that the X-ray spectrum of $\lambda$~Cep is not sensitive to the carbon abundance, as the strongest \ion{C}{v} and \ion{C}{vi} lines are located at wavelengths beyond 30\,\AA\ and fall thus in the heavily absorbed part. The same holds for the absorption edges of different ionization stages of carbon present in the cool wind. Carbon has therefore no measurable impact on the X-ray spectrum of $\lambda$~Cep. However, the RGS spectrum is quite sensitive to oxygen. If we let the abundance of oxygen vary in the fit, we obtain an abundance slightly higher than solar, but fully consistent with the solar abundance within the errors. In Table\,\ref{xspecfits}, we present results with an oxygen abundance set to solar.

From the absorption-corrected X-ray fluxes in Table\,\ref{xspecfits} and a distance of 0.95\,kpc (as adopted by Bouret et al.\ \cite{Bouret}\footnote{This is significantly larger than the {\it Hipparcos} distance of ($606 \pm 81$)\,pc (van Leeuwen \cite{Hipparcos}).}), one can estimate an X-ray luminosity of $1.9 \times 10^{32}$\,erg\,s$^{-1}$ in the 0.5 -- 10\,keV energy band. Bouret et al.\ (\cite{Bouret}) infer a bolometric luminosity of $\log{\frac{L_{\rm bol}}{L_{\odot}}} = 5.80$. This in turn leads to $\log{\frac{L_{\rm X}}{L_{\rm bol}}} = -7.10 \pm 0.02$ for $\lambda$~Cep. Note that the quoted error only accounts for the dispersion in the value of the absorption-corrected X-ray fluxes. It does not account for any uncertainties on the bolometric luminosity or the stellar distance. 
This $\frac{L_{\rm X}}{L_{\rm bol}}$ ratio is towards the fainter end of, but still compatible with, the average relation for O-stars observed with {\it XMM-Newton} as reported by Naz\'e (\cite{YN}). Therefore, the level of X-ray emission of $\lambda$~Cep appears rather normal for a single O-star, and there is no overluminosity due to a magnetically confined wind, contrary to what is seen in a sample of OB stars with magnetic detections (Naz\'e et al.\ \cite{NazeBfield}).

\subsection{Individual line profiles \label{indivprof}}
\begin{figure}
\resizebox{8cm}{!}{\includegraphics{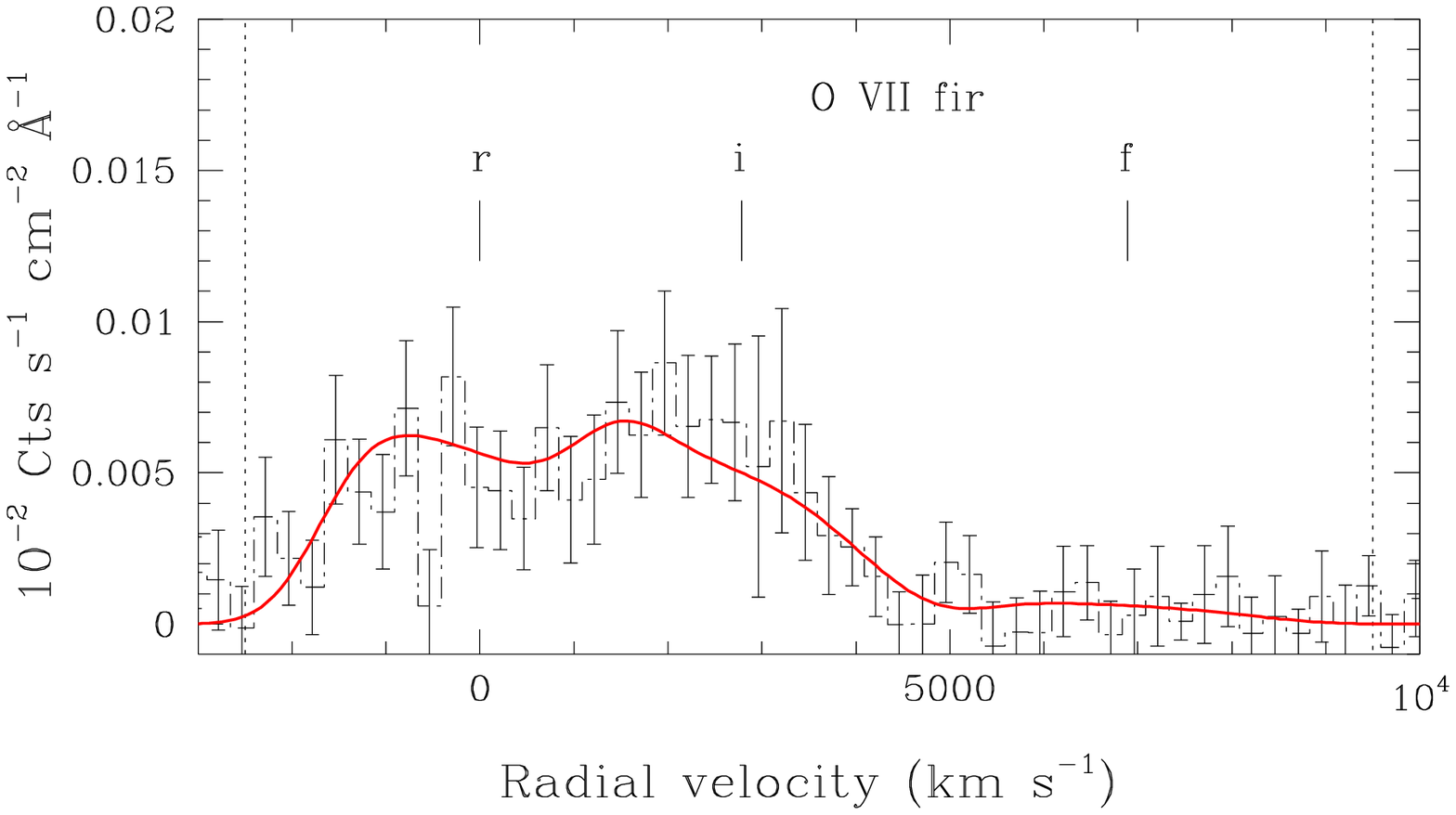}}
\resizebox{8cm}{!}{\includegraphics{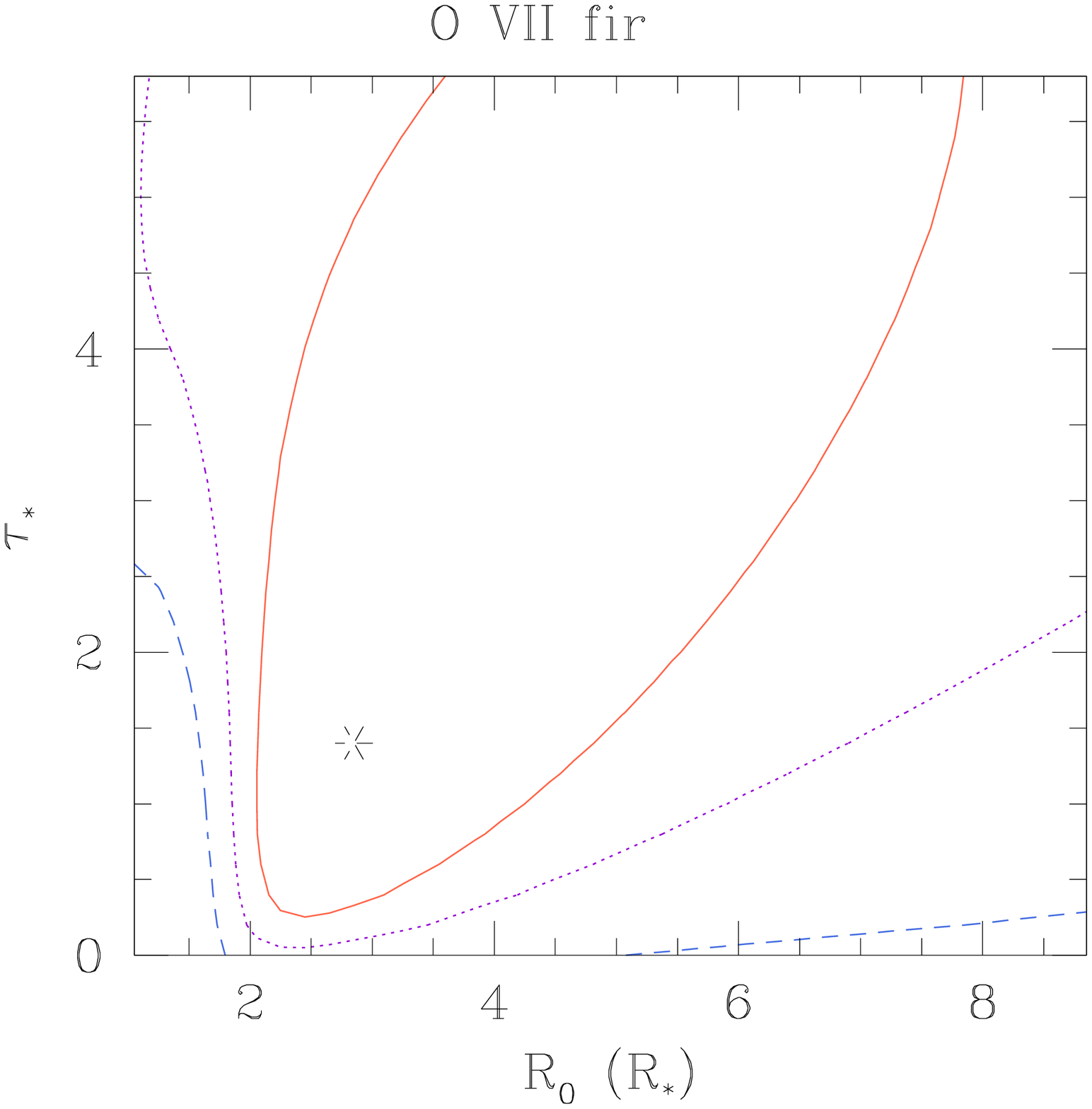}}
\caption{Same as Fig.\,\ref{fits}, but for the He-like triplet of \ion{O}{vii}.\label{fitsOVII}}
\end{figure}

\begin{figure}
\resizebox{8cm}{!}{\includegraphics{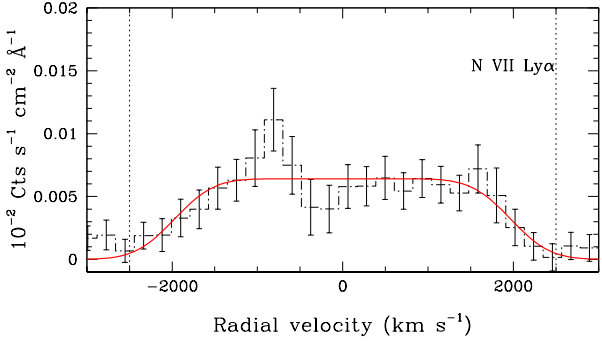}}
\caption{Best fit for the \ion{N}{vii} Ly$\alpha$ line, assuming no contamination by \ion{N}{vi} $\lambda$\,24.9.\label{NLya}}
\end{figure}

As a next step in the analysis of the RGS spectrum of $\lambda$~Cep, we have performed fits of individual spectral lines. If the X-ray emission arises from a hot plasma distributed throughout the otherwise cool wind and moving along with the latter, we can expect the spectral lines to be broad and skewed (MacFarlane et al.\ \cite{MacFarlane}, Owocki \& Cohen \cite{OC1,OC2}). Fitting the lines with such a model can provide us with a first order estimate of where the X-ray plasma is located\footnote{Although we caution that only a fit of the full spectrum consistently deals with the blends and contributions from plasmas at different temperatures and positions in the wind, see Herv\'e et al.\ (\cite{Herve}) and Sect.\,\ref{global}.}, and most of all, it can show us whether or not the lines are indeed broad and skewed. Given the limited quality of our data, we have fitted the profiles with the simplest possible model. Previous studies (Sundqvist et al.\ \cite{Sundqvist}, Herv\'e et al.\ \cite{Herve2}) have shown that, unless the porosity lengths are very large, porosity has a rather marginal effect on X-ray line profiles. Large porosity lengths are ruled out by the observed line profiles (Herv\'e et al.\ \cite{Herve}, Leutenegger et al.\ \cite{Leutenegger13}). In our calculations, we thus assume that there is no porosity. We further assume that the plasma filling factor has no radial dependence. The velocity of both the hot plasma and the cool wind is taken as $v(r) = v_{\infty} \left(1 - \frac{R_*}{r}\right)$ with $v_{\infty} = 2100$\,km\,s$^{-1}$ (Bouret et al.\ \cite{Bouret}). The only remaining free parameters are the inner radius of the emission region $R_0$ and the typical optical depth of the wind $\tau_* = \frac{\kappa\,\dot{M}}{4\,\pi\,R_*\,v_{\infty}}$. We have fitted the profiles using our own code where the synthetic profiles are convolved with the line-spread function of the RGS (FWHM = 0.07\,\AA) and are then compared to the observed profile. As a consistency check, we have also used the {\tt windprofile} models\footnote{http://heasarc.gsfc.nasa.gov/xanadu/xspec/models/windprof.html} implemented under {\tt xspec}. Both methods give generally consistent results. The results of some of our fits are illustrated in Figs.\,\ref{fits} - \ref{NLya}.

One of the strongest lines in the RGS spectrum of $\lambda$~Cep is the \ion{Fe}{xvii} blend near 15\,\AA. Whilst this complex is dominated by the blend of the \ion{Fe}{xvii} $\lambda\lambda$\,15.014, 15.265 lines, other transitions, such as \ion{O}{viii} Ly$\gamma$ at 15.175\,\AA\ probably also contribute. This makes this feature not well suited for line profile fitting. The same restriction applies to the \ion{Ne}{ix} He-like triplet which is blended with \ion{Fe}{xvii} lines. 

The \ion{Fe}{xvii} lines around 17\,\AA\ are in principle better candidates for line profile fitting. The \ion{Fe}{xvii} $\lambda$\,16.78 line is relatively well isolated. On the other hand, \ion{Fe}{xvii} $\lambda\lambda$\,17.051, 17.096 are heavily blended. The latter of these lines arises from a metastable level that can be depleted in the presence of a strong photospheric UV radiation field (Mauche et al.\ \cite{Mauche}, Herv\'e et al.\ \cite{Herve}). As a result, the intensity of the $\lambda$\,17.096 line relative to $\lambda$\,17.051 is difficult to predict, making this blend more problematic for line profile fitting. Comparing the ratio between the observed fluxes of the \ion{Fe}{xvii} $\lambda\lambda$\,17.051, 17.096 blend and the \ion{Fe}{xvii} $\lambda$\,16.78 line to the corresponding ratio of intrinsic emissivities of these features, we empirically estimate that the $\lambda$\,17.096 line has its contribution reduced by at least a factor two with respect to its theoretical emissivity. Fig.\,\ref{fits} illustrates the fits obtained under this assumption.  

The three components of the \ion{O}{vii} He-like triplet were fitted simultaneously assuming identical line parameters ($\tau_*$, $R_0$) for all three lines. As can be seen on Fig.\,\ref{fitsOVII}, the forbidden line is strongly suppressed with a 1$\sigma$ upper limit on the ${\cal R} = f/i$ ratio of 0.12. Accounting for the UV pumping by a diluted photospheric radiation field, the upper limit on ${\cal R}$ translates into an upper limit on $R_0$ of 12.5\,R$_*$, which is entirely consistent with the best-fit $R_0 = 2.85$\,R$_*$ derived from the fit of the line profile.  

The situation is more complicated in the case of the \ion{N}{vii} Ly$\alpha$ line which is blended with \ion{N}{vi} $\lambda$\,24.9. If we assume that both lines arise in the same region of the wind, we can estimate the contribution of the \ion{N}{vi} $\lambda$\,24.9 line to the blend from the temperature dependence of the emissivities derived in our global fit of the broadband spectra (Sect.\,\ref{broadband}) or in the global fit of the RGS spectrum (see the forthcoming Sect.\,\ref{global}). In the former case, the combined EPIC and RGS spectra were fitted with plasma components at temperatures of 0.29 and 0.86\,keV. At these temperatures, the contribution of the \ion{N}{vi} line is expected to be respectively $\sim 60$ and 400 times lower than that of the Ly$\alpha$ line. If we consider instead the plasma temperatures found in our global fit of the RGS spectrum (Sect.\,\ref{global}), we expect the strongest contribution of the \ion{N}{vi} line to arise from the 0.20\,keV component. However, even in this case, the emissivity of the \ion{N}{vii} Ly$\alpha$ doublet exceeds that of the \ion{N}{vi} $\lambda$\,24.9 by a factor $\sim 20$. In the case of $\lambda$~Cep, the \ion{N}{vi} $\lambda$\,24.9 line should thus play a negligible role, suggesting that we can treat this blend as consisting only of the \ion{N}{vii} Ly$\alpha$ doublet. The results obtained under this assumption are shown in Fig.\,\ref{NLya}. As one can see on this figure, the fit is of quite limited quality, especially in the blue part. The best-fit parameters differ significantly from those of the other lines. The best-fit wind optical depth is now 0.0 and the best-fit inner radius of the emitting region is no longer constrained, except that it must be larger than 8.5\,R$_*$.  

\begin{figure}
\resizebox{8cm}{!}{\includegraphics{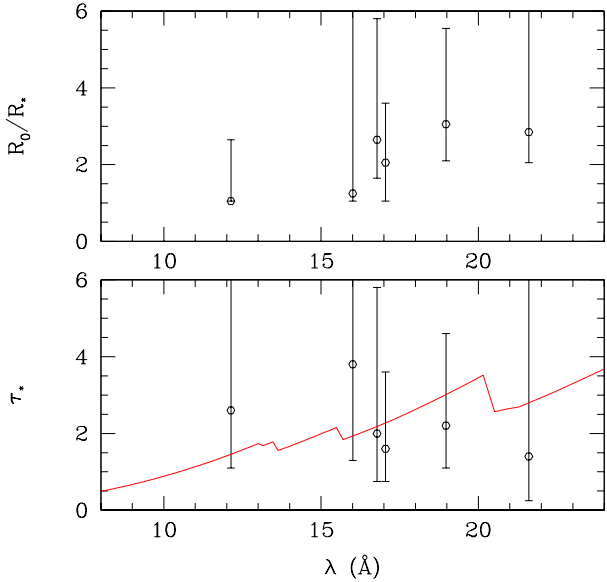}}
\caption{Parameters of the best fits for those lines that could be fitted. The continuous line in the lower panel illustrates the typical optical depth as computed from the CMFGEN model used by Bouret et al.\ (\cite{Bouret}) and adopting an opacity $\kappa$ evaluated at a distance of 2\,R$_*$ from the centre of the star.\label{opacity}}
\end{figure}

To improve the fit of the blue part of the profile and simultaneously bring the fit parameters into better agreement with the trends seen in the other lines, we would need to assume a significantly larger contribution of the \ion{N}{vi} $\lambda$\,24.9 line. As pointed out above, the known temperature components do not allow for such a contribution. Yet, it could arise from a significantly cooler plasma component that would otherwise be hidden by the large interstellar absorption that prevents us from seeing the \ion{N}{vi} He-like triplet emission. In the case of $\zeta$~Pup, Herv\'e et al.\ (\cite{Herve}) found that a fit of the RGS spectrum required a four-temperature plasma, with the coolest component at 0.10\,keV accounting for the bulk of the \ion{N}{vi} He-like triplet emission. At this plasma temperature, the contribution of \ion{N}{vi} $\lambda$\,24.9 would be about half that of the \ion{N}{vii} Ly$\alpha$ doublet. To investigate the possibility that such a cool component might also exist in the wind of $\lambda$~Cep, we have repeated the fits of the \ion{N}{vii} Ly$\alpha$ + \ion{N}{vi} $\lambda$\,24.9 blend assuming a relative contribution of the \ion{N}{vi} feature varying from 0.0 to 0.9 times the contribution of \ion{N}{vii}. Increasing the relative contribution of \ion{N}{vi} leads to an increase of $\tau_*$ as expected, but $R_0$ remains at the upper limit of our parameter grid and the fit quality actually degrades. Hence, we conclude that the present model cannot easily explain the morphology of this line unless its parameters deviate significantly from those of other lines. \\

\begin{table}[h]
\caption{Best-fit parameters for the lines where a good fit quality was achieved.\label{fitlines}}
\begin{center}
\begin{tabular}{l c c}
\hline
\multicolumn{1}{c}{Line} & $R_0$ (R$_*$) & $\tau_*$ \\
\hline
\vspace*{-3mm}\\
\ion{Ne}{x} $\lambda$\,12.134 & $1.05_{-0.00}^{+1.60}$ & $2.60_{-1.50}^{+...}$ \\
\vspace*{-3mm}\\
\ion{O}{viii} $\lambda$\,16.006 & $1.25_{-0.20}^{+...}$ & $3.80_{-2.50}^{+...}$ \\
\vspace*{-3mm}\\
\ion{Fe}{xvii} $\lambda$\,16.780 & $2.65_{-1.00}^{+3.15}$ & $2.00_{-1.25}^{+3.80}$ \\
\vspace*{-3mm}\\
\ion{Fe}{xvii} $\lambda\lambda$\,17.051-17.096 & $2.05_{-1.00}^{+1.55}$ & $1.60_{-0.85}^{+2.00}$ \\
\vspace*{-3mm}\\
\ion{O}{viii} $\lambda$\,18.970 & $3.05_{-0.95}^{+2.50}$ & $2.20_{-1.10}^{+2.40}$ \\
\vspace*{-3mm}\\
\ion{O}{vii} $\lambda\lambda$\,21.602-22.098 & $2.85_{-0.80}^{+4.90}$ & $1.40_{-1.15}^{+...}$ \\
\vspace*{-3mm}\\
\ion{N}{vii} $\lambda$\,24.780 & $8.85^{+...}$ & $0.00_{-0.00}$ \\
\vspace*{-3mm}\\
\hline
\end{tabular}
\end{center}
\tablefoot{$R_0$ is measured from the centre of the star. Whenever possible, the 1$\sigma$ errors are estimated from the contours in the parameter space. A `...' indicates situations where the 1$\sigma$ contours extend beyond the range of parameters explored in our calculations.}
\end{table}

Table\,\ref{fitlines} summarizes the results of our fits and Fig.\,\ref{opacity} illustrates the overall behaviour of the best-fit $\tau_*$ and $R_0$ values as a function of wavelength for those lines where a good fit was achieved. Most transitions have best-fit $R_0$ values consistent with an emission region starting between 2 and 3\,R$_*$. The only exceptions are the shortest wavelength transitions that seem to originate very near the photosphere, although, also in these cases, the 1$\sigma$ confidence ranges are still consistent with an onset radius between 2 and 3\,R$_*$ from the centre of the star. There is no obvious trend of $\tau_*$ with wavelength, though we note that the (large) error bars overlap with the expected $\tau_*(\lambda)$ relation at a distance of 2\,R$_*$ from the centre of the star assuming the mass-loss rate inferred by Bouret et al.\ (\cite{Bouret}). 

\subsection{Global fit of the RGS spectrum \label{global}}
Under the assumption of a wind embedded plasma, the X-ray spectrum of a single non-magnetic O-type star stems from the combination of emissions arising from plasma at different temperatures and different locations in the wind. Hence, the above approach of fitting individual line profiles only provides a first order approximation of the plasma properties. A more rigorous approach is therefore needed. Herv\'e et al.\ (\cite{Herve2,Herve,Herve3}) have developed a code that aims at achieving a consistent fit of the high-resolution X-ray spectrum, simultaneously accounting for the presence of several plasma components with different locations inside the wind. In this section, we apply this code to the specific case of the RGS spectrum of $\lambda$~Cep. 

In our model, the part of the wind that contains the X-ray plasma is divided into a number of spherically symmetric shells. To allow for a position-dependent filling factor\footnote{The filling factor is defined here as the ratio of the X-ray emitting volume and the total volume of the shell under consideration.} of the hot plasma, we adopt shells with a minimum width of 0.1\,R$_*$ (see also below). The X-ray emissivities of each of these shells are computed assuming an optically thin thermal plasma (using the AtomDB code, see Foster et al.\ \cite{foster12} and references therein). The important parameters are the plasma temperature and the quantity of hot gas contained in the shell. The shells are assumed to move along with the ambient cool wind and their emission is Doppler-shifted into the frame of reference of the observer. The optical depth of the cool wind material is computed along each line of sight from the observer to an emitting shell using the $\kappa(\lambda,r)$ relations computed with CMFGEN (see Herv\'e et al.\ \cite{Herve2,Herve,Herve3} for details).

The UV pumping of the 2\,$^3S_1$ level in He-like ions to their 2\,$^3P_{1,2}$ level and the resulting enhancement of the intercombination line at the expense of the forbidden line are taken into account following the approach of Leutenegger et al.\ (\cite{Leutenegger06}). For this purpose, we adopt the UV fluxes from our CMFGEN model. 
 
The RGS spectrum features lines that arise from plasmas at different temperatures (e.g.\ \ion{Fe}{xvii} lines are preferentially produced by a hot plasma with $kT$ close to 0.6 - 0.7\,keV, whereas the emissivity of \ion{N}{vii} lines peaks around 0.2\,keV). To reproduce the observed spectrum, we therefore need to combine models with different temperatures and with different radial dependencies of their filling factors. We thus need to find a combination of our models that produces a synthetic spectrum that best matches our observed spectrum. To this aim, we have to build a database which samples the different parameters of the synthetic spectra simulator code such as the inner and outer boundary of the emission region, the abundances of CNO, the mass-loss rate, the wind velocity, the porosity length and, of course, the X-ray emitting plasma temperature. In the present case, several parameters were frozen. Indeed, we adopted the wind velocity law of Bouret et al.\ (\cite{Bouret}) and we neglected porosity (see Sect.\,\ref{indivprof}). Solar CNO abundances did not allow to fit the data, and we thus adopted the same chemical composition as derived by Bouret et al.\ (\cite{Bouret}). Our fitting routine, based on the Bound Variable Least Square algorithm, systematically explores this database, combining up to four plasma components with the same values for the chemical composition and mass-loss rate. 

\begin{figure}[htb]
\resizebox{8cm}{!}{\includegraphics{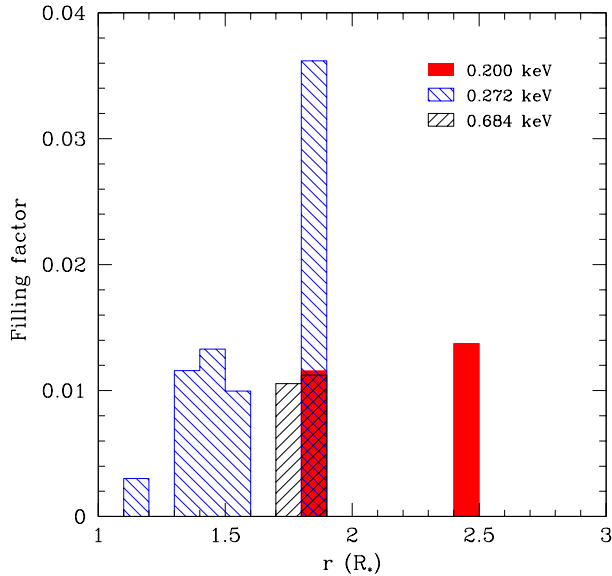}}
\caption{Radial distribution of the filling factor of the hot plasma inside the stellar wind of $\lambda$~Cep as inferred from our fit of the RGS spectrum shown in Fig.\,\ref{RGSlCep}.\label{spatial}}
\end{figure}

\begin{figure}[htb]
\resizebox{8cm}{!}{\includegraphics{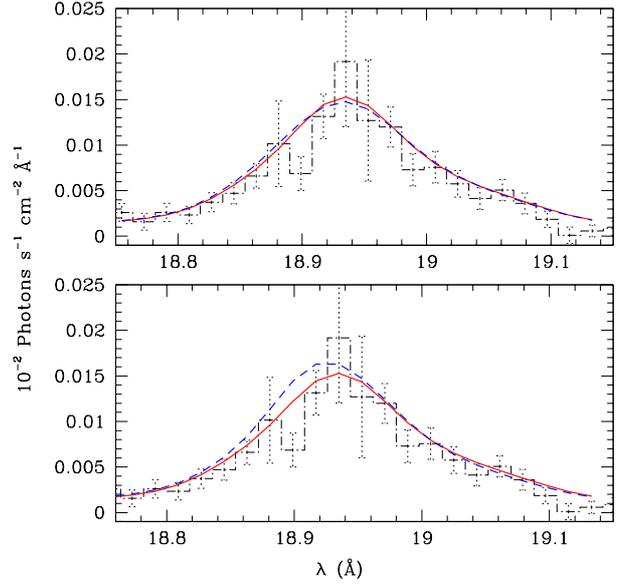}}
\caption{Impact of some model assumptions on the fits of the RGS spectrum of $\lambda$~Cep, illustrated by the example of the O\,{\sc viii} Ly$\alpha$ line. In the top panel, the red solid line illustrates the best-fit model (assuming minimum shell width of 0.1\,R$_*$), whilst the blue dashed line corresponds to the best fit for a minimum shell width of 1\,R$_*$. The histogram yields the observed line profile. The bottom panel shows the same figure, except for the blue dashed line which corresponds now to the best fit where we have artificially set the onset radius of the X-ray emission region to be located at $r \geq 2$\,R$_*$.\label{tests}}
\end{figure}

Overall our best-fit model does a good job in reproducing the observed RGS spectrum (see Fig.\,\ref{RGSlCep}). Still it fails to reproduce some features. For instance, whilst the strong \ion{Fe}{xvii} $\lambda$\,15.014 line is very well reproduced, the strengths of the \ion{Fe}{xvii} $\lambda\lambda$\,15.265, 16.780, 17.051 lines are overpredicted by our model. For the \ion{Fe}{xvii} $\lambda$\,17.051 line, we found that the contribution of \ion{Fe}{xvii} $\lambda$\,17.096 must be essentially negligible as a result of UV pumping of the upper level of this transition.   

The best-fit mass-loss rate of $\lambda$~Cep was found to be $1.5 \times 10^{-6}$\,M$_{\odot}$\,yr$^{-1}$, in excellent agreement with previous studies ($1.5 \times 10^{-6}$\,M$_{\odot}$\,yr$^{-1}$, Sundqvist et al.\ \cite{Sund1,Sund2}; $1.4 \times 10^{-6}$\,M$_{\odot}$\,yr$^{-1}$, Bouret et al.\ \cite{Bouret}; $1.6 \times 10^{-6}$\,M$_{\odot}$\,yr$^{-1}$, \v{S}urlan et al.\ \cite{Surlan}) that deal with the clumpiness of the stellar wind in different ways\footnote{Sundqvist et al.\ (\cite{Sund1,Sund2}) and \v{S}urlan et al.\ (\cite{Surlan}) independently analysed UV wind resonance lines accounting for the effects of clumps of any optical depth (from optically thin to optically thick). Sundqvist et al.\ (\cite{Sund1,Sund2}) also account for porosity in velocity space, whilst \v{S}urlan et al.\ (\cite{Surlan}) considered the velocity dispersion inside the clumps. Bouret et al.\ (\cite{Bouret}) simultaneously fitted the FUV, UV and optical spectra adopting a parametric filling-factor approach and assuming optically thin clumps.}. This agreement indicates that high-resolution X-ray spectra can indeed be used to constrain the mass-loss rates of O-type stars. The above values of $\dot{M}$ are lower than the theoretical mass-loss rate based on the recipe of Vink et al.\ (\cite{Vink2}). Indeed, applying the latter recipe to the stellar parameters as inferred by Bouret et al.\ (\cite{Bouret}) yields $4.5 \times 10^{-6}$\,M$_{\odot}$\,yr$^{-1}$, i.e.\ a factor 3 larger than what we have found here. Various studies comparing mass-loss rates derived from observations with those predicted by the Vink et al.\ (\cite{Vink2}) formalism have revealed similar discrepancies (e.g.\ Bouret et al.\ \cite{Bouret}, Puls et al.\ \cite{Puls}), with the Vink et al.\ (\cite{Vink2}) rates being a factor 2 -- 3 too large. Our results indicate that $\lambda$~Cep is yet another case where this conclusion applies.

Three plasma components were needed to achieve the fit shown in Fig.\,\ref{RGSlCep}. The best-fit plasma temperatures were 0.200, 0.272 and 0.684\,keV. The radial distributions of their filling factors are shown in Fig.\,\ref{spatial}. Although we caution that this distribution is likely not unique (see below), it is worth pointing out that the bulk of the hot plasma is located within less than 2\,R$_*$ above the stellar surface and some emission can actually arise very close to the stellar surface. The same conclusion is reached for models which have slightly different temperatures but produce fits of almost equal quality. This result also agrees quite well with the $R_0$ values found in Fig.\,\ref{opacity}. 

Although our algorithm systematically explores the parameter space to search for the best combination of a predefined number of plasma components, we caution that the distribution in Fig.\,\ref{spatial} is likely not unique. For instance, while we found that two plasma temperatures are clearly not sufficient to reproduce the observed spectrum, the possible presence of a fourth lower-temperature component cannot be constrained. Moreover, the limited S/N ratio of the RGS spectrum of $\lambda$~Cep and the strong interstellar absorption prevent us from gathering stronger constraints on the temperatures and locations of the emitting plasma that would lift some degeneracies on the radial distribution of the filling factor. The latter  might also depend on some of our assumptions, such as the minimum width of the emitting shells. Indeed, Fig.\,\ref{spatial} shows that the various emission regions extend only over a few times the minimum shell width. To quantify the sensitivity of our fit on the assumed minimum width of the shells, we have performed another fit setting the minimum shell width artificially to 1\,R$_*$. The top panel of Fig.\,\ref{tests} illustrates the comparison between the best fit obtained under this assumption and the one of Fig.\,\ref{RGSlCep}. To enhance the visibility of this comparison, we focus on a single line (O\,{\sc viii} Ly$\alpha$), although we stress that the full spectrum was fitted. As can be seen, except for a marginal change in line width, there is essentially no difference between the line profiles obtained for both assumptions. This is actually not surprising as Fig.\,\ref{spatial} reveals shells of the dominant plasma component (the one at 0.272\,keV) to be present between $r = 1.1$ and $r = 1.9$\,R$_*$, i.e.\ over a range of almost 1\,R$_*$. We thus conclude that the detailed radial distribution of the hot plasma filling factor depends on the assumptions that we make on the shell width.

How robust is then the finding that the X-ray emission starts already at 1.1\,R$_*$? To check this, we have done yet another fit where we artificially require the onset radius of the X-ray emission region to be located at $r \geq 2$\,R$_*$. The result is shown in the bottom panel of Fig.\,\ref{tests}. This time, we see that a larger onset radius leads to a blue-shift of the line centroid. Whilst this effect is small if we consider a single line (as shown in Fig.\,\ref{tests}), considering the spectrum as a whole leads to a better fit for the onset radius at 1.1\,R$_*$. 

At first sight, the onset of the X-ray emission that close to the stellar surface is a challenge for theoretical models. Whilst the hydrodynamical simulations of Feldmeier et al.\ (\cite{FPP}) predict that the instability of the line driving mechanism already produces considerable structures as close as 0.1\,R$_*$ above the photosphere, in these models, the bulk of the X-ray emission arises between 1 and 3\,R$_*$ above the photosphere. More recently, Sundqvist \& Owocki (\cite{SO}) included the effect of limb-darkening in their hydrodynamical simulations. These authors noted that limb darkening favours the development of wind structures already at $r = 1.1$\,R$_*$, although these simulations provide no information about the level of X-ray emission. Such an early onset of the X-ray emission could also affect the formation of important UV resonance lines as well as optical lines formed at the base of the wind. From the observational point of view, we note that similar situations have been reported for other stars using independent methods. For instance, in their analysis of the O9.5\,II star $\delta$~Ori, Shenar et al.\ (\cite{Shenar}) derived an onset radius of $\sim 1.1\,R_*$.   

Finally, one may wonder about the significance of the outer boundary of the X-ray emitting plasma near 2.5\,R$_*$. In some models, we found indeed plasma components that are present out to much larger radii. However, in these cases, the corresponding filling factors were very low so that these regions contributed only a marginal fraction of the line flux. Leutenegger et al.\ (\cite{Leutenegger13}) showed that the existence of an upper radial cutoff impacts the shape of the lines (their Fig.\,7), leading to a more asymmetric profile for a given value of $\tau_*$. For our analysis of Sect.\,\ref{indivprof}, where we made the assumption of an infinite emission region, this may imply an overestimate of the values of $\tau_*$.

In summary, we thus conclude that the results that the X-ray emission of $\lambda$~Cep starts already at 1.1\,R$_*$ and arises from a region that spans about 2\,R$_*$ in the wind are quite robust.

This distribution is actually quite different from the one obtained in the case of $\zeta$~Pup (Herv\'e et al.\ \cite{Herve}). In the latter star, the softer emission starts at 0.5\,R$_*$ above the photosphere and reaches out to very large distances in the wind. The harder emission component in the spectrum of $\zeta$\,Pup arises from between 2 and 3\,R$_*$ above the photosphere. We thus see that the X-ray emitting plasma is much closer to the stellar surface in the case of $\lambda$~Cep and covers a narrower range of distances than in the wind of $\zeta$~Pup. 
 
\section{Conclusion \label{sectdisc}}
In this paper, we have presented the first-ever X-ray analysis of the O6\,Ief star $\lambda$~Cep. Because of its spectral type, this star is certainly the closest cousin of $\zeta$~Pup, although, because of its larger distance and higher interstellar absorption, its X-ray flux at Earth is much lower. We have further gathered coordinated optical spectroscopy to monitor the variability of $\lambda$~Cep during our four {\it XMM-Newton} observations. These data led to the following results:
\begin{itemize}
\item Optical spectroscopy revealed variability of the H$\alpha$ and \ion{He}{ii} $\lambda$\,4686 emission lines. Whilst the former appears to vary on a timescale of 4.1\,days, very likely associated with the rotation of the star, the latter varies on an unrelated timescale (18.3\,hrs). Given the proximity of the formation regions of both lines, as derived from 1-D model atmosphere codes, this difference is a puzzle that needs to be addressed in future studies. 
\item While there is only marginal X-ray variability during individual X-ray pointings, we find a significant (5$\sigma$) inter-pointing variability. Although the timescale of these variations cannot be constrained by the sole X-ray data, it turns out to be compatible with the 4.1\,days timescale derived from the H$\alpha$ data. One of the  maxima of the X-ray emission seems to occur when the EW of H$\alpha$ is minimum, indicating a phase shift between the two variations. 
\item The high-resolution X-ray spectrum of $\lambda$~Cep reveals broad and slightly skewed emission lines. As in the case of $\zeta$~Pup, most of these lines are rather well reproduced by a model that assumes to first order a spherically symmetric distribution of shock-heated plasma embedded in the cold wind and moving along with the latter. 
\item A global fit of the RGS spectrum indicates that the bulk of the hot plasma must be located close to the stellar surface and yields a mass-loss rate in very good agreement with previous determinations based on optical and UV spectra.
\end{itemize}
  
At first sight, a rotational modulation of the X-ray flux could hint at the effect of wind-confinement by a moderately strong dipolar magnetic field (Babel \& Montmerle \cite{BM}, ud-Doula \& Owocki \cite{udDoula}) which would force the wind, out to the Alfv\'en radius, into rigid co-rotation with the star. However, such a global magnetic field has not been detected (e.g.\ David-Uraz et al.\ \cite{DU}). Furthermore, if this scenario applies in the case of Oef stars, the magnetically confined wind does not extend over a wide part of the X-ray emission region. Otherwise, one would expect to observe a strong X-ray overluminosity (Naz\'e et al.\ \cite{NazeBfield}) along with some narrow, symmetric lines, as seen notably for $\theta^1$~Ori\,C (Gagn\'e et al.\ \cite{Gagne}) and HD~148\,937 (Naz\'e et al.\ \cite{HD148937}). However, this is definitely not the case neither for $\zeta$~Pup (Naz\'e et al.\ \cite{zetaPup1}, Herv\'e et al.\ \cite{Herve}), nor for $\lambda$~Cep, as we have shown here. 
\begin{figure}[htb]
\resizebox{8cm}{!}{\includegraphics{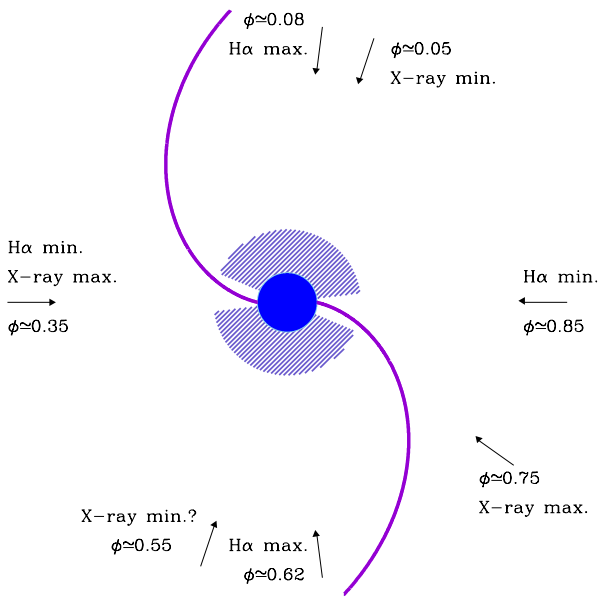}}
\caption{Cartoon of the CIRs in the wind of $\lambda$~Cep and its impact on the X-ray and H$\alpha$ emission. The H$\alpha$ emission region is shown by the blue hatched area. Outside the CIR, the X-ray emission region overlaps with the H$\alpha$ emission region. Near the base of the CIR, excess X-ray emission is produced and H$\alpha$ emission is reduced.\label{cartoon}}
\end{figure}

Henrichs \& Sudnik (\cite{HenSud2}) argue that localized magnetic fields might exist in the case of $\lambda$~Cep, leading to spots corotating with the star that might be at the origin of corotating interaction regions (Cranmer \& Owocki \cite{CO}, Lobel \& Blomme \cite{LB}). Harries (\cite{Harries}) performed Monte-Carlo line profile calculations for a single corotating spiral density enhancement in the wind of an O supergiant. He showed that such a one-armed density wave actually produces a double-wave modulation of the equivalent width similar to what we have observed for the H$\alpha$ line of $\lambda$~Cep. In the simulated line profiles, minimum EW occurs at phases at which the base of the spiral is eclipsed by the star or crosses in front of the latter (Harries \cite{Harries}). Of course, such double-wave modulations could also arise if there were two spots. Indeed, Lobel \& Blomme (\cite{LB}) performed 3-D radiative transfer calculations with hydrodynamical models to simulate the discrete absorption components due to CIRs in the B0.5\,Ib star HD~64\,760. These authors tested models with one or two spots (located on opposite sides of the stellar equator) and found a slightly better agreement between their simulations and the observed variations for the two-spot model\footnote{Note that in this specific case, unlike what we suggest for $\lambda$~Cep, the spots do not seem to corotate with the star, but form in a dynamical structure at the base of the wind that trails behind the stellar rotation (Lobel \& Blomme \cite{LB}).}.    

The question that arises is whether the same CIR could impact the level of the observed X-ray emission. Our results suggest that this is the case. Such a modulation could be either a consequence of extra emission produced in the CIR itself or of the associated density enhancement which would lead to a modulation of the column-density towards the hot plasma. A candidate for the first scenario could be X-ray emission produced behind the shock associated with the CIR. Heating the plasma to $kT = 0.200$ via this mechanism requires a velocity jump of 400\,km\,s$^{-1}$. Such values were deemed possible  in the work of Cranmer \& Owocki (\cite{CO}). However, at least for HD~64\,760, the more recent hydrodynamical simulations of Lobel \& Blomme (\cite{LB}) yield maximum velocity differences across the CIR that do not exceed 140\,km\,s$^{-1}$. Such values would not be sufficient to provide a significant X-ray emission. An alternative scenario to produce the additional X-ray emission could be the action of the localized magnetic fields that were proposed to produce the spots that generate the CIRs (Henrichs \& Sudnik \cite{HenSud2}). An excess of magnetically-heated plasma near the base of the CIR could alter the local ionization, thereby creating a hole in the H$\alpha$ emission region which could explain the apparent anti-correlation between the X-ray and H$\alpha$ emissions. 

Figure\,\ref{cartoon} illustrates a possible configuration. The shape of the CIR in Fig.\,\ref{cartoon} is based on the expression of the spiral streakline of Fullerton et al.\ (\cite{Full}). In Sect.\,\ref{global} we found that the bulk of the X-ray emission of $\lambda$~Cep arises from the inner parts of the wind. Therefore, in view of Fig.\,\ref{cartoon}, it appears unlikely that the X-ray modulation could be due to absorption by the density enhancement associated with the curved part of the CIR. Indeed, maximum X-ray emission is observed when the base of the CIR is roughly aligned with the line of sight and the projection of the curved part of the CIR on the plane perpendicular to the line of sight covers a wide area. A more plausible explanation would thus be that the base of the CIR produces itself some X-ray emission. 

Considering the results of Naz\'e et al.\ (\cite{zetaPup2}) on $\zeta$~Pup, Massa et al.\ (\cite{Massa}) on $\xi$~Per, and our own results on $\lambda$~Cep, observational evidence is accumulating for the impact of large-scale wind structures on the X-ray emission of O-type stars even though current theoretical models have difficulties to explain this. Because such structures might be transient, coordinated optical and X-ray observations are mandatory to firmly establish this link. Such studies are pushing the {\it XMM-Newton} and {\it Chandra} observatories to the limits of their capabilities, but are clearly worth the effort. In the more distant future, these studies will greatly benefit from the unprecedented collecting area of ESA's {\it Athena} X-ray observatory which will allow us to observe the impact of the structures on individual X-ray emission lines (see Sciortino et al.\ \cite{Sciortino}). 
\section*{Acknowledgements}
We thank the referee, Dr.\ Jo Puls, for a very constructive and helpful report.The Li\`ege team acknowledges support from the Fonds de Recherche Scientifique (FRS/FNRS), through an ARC grant for Concerted Research Actions, financed by the Federation Wallonia-Brussels, through the XMM and Gaia-DPAC PRODEX contract as well as by the Communaut\'e Fran\c caise de Belgique. The TIGRE facility is funded and operated by the universities of Hamburg, Guanajuato and Li\`ege. AH acknowledgeds financial support from the Agence Nationale de la Recherche (France) through grant ANR-11-JS56-0007. We are grateful to Prof.\ John Hillier for making the CMFGEN atmosphere code available and for his assistance to the users.
 
\end{document}